\documentclass[a4paper,11pt]{article}
\usepackage{jheppub} 
\graphicspath{{Figs/}}

\usepackage{placeins}
\usepackage{physics}
\usepackage{xstring}
\usepackage{tabularx}
\usepackage{longtable}
\usepackage[percent]{overpic}
\usepackage{tikz}
\newcommand{\pdag}{^{\phantom{\dagger}}}
\renewcommand{\dag}{^{\dagger}}
\newcommand{\cbf}[1]{\hat{\bm{c}}\pdag_{#1}}
\newcommand{\cdagbf}[1]{\hat{\bm{c}}\dag_{#1}}

\newcommand{\cfig}[1]{Fig.~\ref{fig:#1}}
\newcommand{\red}[1]{{\color{red}{#1}}}
\newcommand{\AD}[1]{#1}
\usepackage{cancel,soul}
\usepackage[normalem]{ulem}
\usepackage{bm}

\usepackage{amsthm}
\theoremstyle{remark}
\newtheorem{remark}{Remark}

\makeatletter
\def\@fpheader{\ }
\makeatother

\title{Conformal Data for the $O(2)$ Wilson-Fisher CFT
in $(2+1)$-Dimensional Spacetime
from Exact Diagonalization and Matrix Product States
on the Fuzzy Sphere}

\author[a, b]{Arjun Dey,}
\emailAdd{arjun.dey@psi.ch}
\author[c]{Loic Herviou,}
\emailAdd{loic.herviou@lpmmc.cnrs.fr}
\author[a, b]{Christopher Mudry,}
\emailAdd{christopher.mudry@psi.ch}
\author[d]{Slava Rychkov}
\emailAdd{slava@ihes.fr}
\author[a, b]{and Andreas Martin L\"auchli}
\emailAdd{andreas.laeuchli@epfl.ch}

\affiliation[a]{
Laboratory for Theoretical and Computational Physics,
PSI Center for Scientific Computing, Theory and Data,
5232 Villigen PSI, Switzerland}
\affiliation[b]{
Institute of Physics,
\'{E}cole Polytechnique F\'{e}d\'{e}rale de Lausanne (EPFL),
1015 Lausanne, Switzerland}
\affiliation[c]{Univ. Grenoble Alpes, CNRS, LPMMC, 38000 Grenoble, France}
\affiliation[d]{Institut des Hautes \'Etudes Scientifiques, 91440 Bures-sur-Yvette, France}

\abstract{
We study at zero temperature a microscopic quantum spin-1 model
on the fuzzy sphere that realizes
the $O(2)$ Wilson-Fisher conformal field theory
(CFT) in $(2+1)$-dimensional spacetime
at a quantum critical point. Here, we use
the fuzzy-sphere regularization as it
preserves the full spatial $SO(3)$ rotational symmetry of the
CFT, enabling the
state-operator correspondence that maps energy eigenstates directly to
CFT operators. Using exact diagonalization (ED) and matrix product
state (MPS) techniques combined with conformal perturbation theory
(CPT), we extract conformal data including scaling dimensions and
operator product expansion (OPE) coefficients. We identify 48 primary
operators and their descendants, organized by the conserved $O(2)$
charge $S^{z}$ and spatial angular momentum $L$.
Our numerical results
for the scaling dimensions of the lowest primary operators show good
agreement with conformal bootstrap predictions. We verify
predictions from the large charge expansion, which provides systematic
predictions for operators carrying large $U(1)$ charge, connecting the
Goldstone mode physics in the ordered phase to phonon primaries at the
critical point.
         }

\begin{document}
\maketitle
\flushbottom

\newpage
\section{Introduction}
\label{sec:intro}

The $O(2)$ conformal field theory (CFT)
in $(2+1)$-dimensional spacetime
describes
the universal behavior at a special type of quantum phase
transition. This transition occurs when a system with an
internal $O(2)$ symmetry
changes from an ordered phase, in which the subgroup
$SO(2)$ is spontaneously broken,
to a disordered phase where the $O(2)$ symmetry is preserved.
The $O(2)$ CFT
in $(2+1)$-dimensional spacetime
is one of the most important examples of a
Wilson-Fisher fixed point~\cite{Wilson_Fisher_Critical_Point_Wilson_1972},
appearing in a wide variety of physical systems.
It describes the critical point
for three-dimensional (two-dimensional) classical (quantum) XY magnets
~\cite{3D_XY_universality_class},
the $\lambda$ transition in liquid helium-4%
~\cite{Hasenbusch_Superfluid_Transition},
and certain quantum phase transitions in models of interacting bosons.

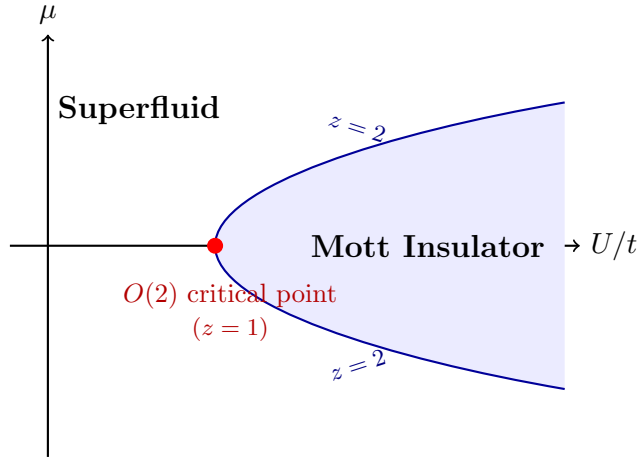
\begin{figure}[t]
    \centering
    \begin{tikzpicture}[scale=1.0]
        \draw[->, thick] (-0.5,0) -- (7,0) node[right] {$U/t$};
        \draw[->, thick] (0,-2.8) -- (0,2.8) node[above] {$\mu$};

        \fill[blue!8]
            (2.2,0)
            .. controls (2.2,0.8) and (4.5,1.5) .. (6.8,1.9)
            -- (6.8,-1.9)
            .. controls (4.5,-1.5) and (2.2,-0.8) .. (2.2,0)
            -- cycle;

        \draw[thick, blue!60!black]
            (6.8,1.9)
            .. controls (4.5,1.5) and (2.2,0.8) .. (2.2,0);
        \draw[thick, blue!60!black]
            (2.2,0)
            .. controls (2.2,-0.8) and (4.5,-1.5) .. (6.8,-1.9);

        \node[font=\large\bfseries] at (5.0, 0) {Mott Insulator};
        \node[font=\large\bfseries] at (1.2, 1.8) {Superfluid};

        \node[blue!50!black, font=\footnotesize, anchor=south, rotate=-18] at (4.0, 1.35) {$z=2$};
        \node[blue!50!black, font=\footnotesize, anchor=north, rotate=18] at (4.0, -1.35) {$z=2$};

        \fill[red] (2.2, 0) circle (3pt);
        \node[below, font=\small, text=red!70!black, align=center] at (2.4, -0.35) {$O(2)$ critical point\\[-1pt]{\footnotesize ($z=1$)}};
    \end{tikzpicture}
    \caption{
Schematic phase diagram of the Bose-Hubbard model with on-site
interaction $U$, hopping amplitude $t$, and chemical potential
$\mu$. The Mott insulating lobe (shaded) is separated from the
surrounding superfluid phase by a quantum phase transition. Along the
generic boundary, the transition has dynamical critical exponent
$z=2$. At the tip of the Mott lobe (red dot), where particle-hole
symmetry is present at integer filling, the transition belongs to the
$O(2)$ universality class in $(2+1)$-dimensional spacetime with $z=1$.}
    \label{fig:phase_diagram_BH}
\end{figure}

A concrete example where this transition appears is the Bose-Hubbard
model~\cite{fisher_PRB_89}, which describes bosons hopping on a square
lattice with hopping amplitude $t$, on-site repulsion $U$, and
chemical potential $\mu$. As shown in \cfig{phase_diagram_BH}, this
model exhibits a quantum phase transition between a Mott insulator and a
superfluid phase. At most points along the phase boundary, the
transition has a dynamical critical exponent $z=2$, meaning time and
space scale differently. However, at the tip of each Mott lobe,
when the system is at integer filling,
a special transition occurs with $z=1$, 
meaning time and space scale the same way.
This tip transition belongs to the $O(2)$ universality class
in $(2+1)$-dimensional spacetime
~\cite{Sachdev_2011}.

The $O(2)$ CFT
in $(2+1)$-dimensional spacetime
has attracted significant theoretical interest,
particularly in connection with the large charge
expansion~\cite{Hellerman15,Monin17,Alvarez-Gaume17,Cuomo:2017vzg,Cuomo2020,Gaume21}. This
approach exploits the fact that when the system is in the superfluid
phase, operators carrying a large amount of the conserved $U(1)$
charge can be understood using
semiclassical methods. The large charge expansion provides systematic
predictions for the scaling dimensions of operators with large charge,
connecting the physics of the superfluid phase to the conformal field
theory at the critical point.

To systematically study the $O(2)$ CFT
in $(2+1)$-dimensional spacetime
and extract its universal
properties, we need a numerical method that preserves
microscopically as many symmetries of the CFT as may be.
The key tool is the state-operator
correspondence~\cite{Rychkov_2017,Di_Francesco_Mathieu_Senechal_1997},
which maps local operators to energy eigenstates on a sphere. A
primary operator with scaling dimension $\Delta$ and angular momentum
$L$ corresponds to an energy eigenstate with energy proportional to
$\Delta/R$ and angular momentum $L$, where $R$ is the radius of the
sphere (working in units where $\hbar=c=1$).
This correspondence works best
when the spatial geometry preserves full rotational symmetry,
because
rotations on the sphere correspond to Lorentz transformations in the CFT.
Although the correspondence is a general property of CFTs,
traditional numerical approaches using periodic boundary conditions
on a torus break continuous rotational symmetry,
making the extraction of CFT data difficult.
Ideally, one would discretize the sphere
while preserving the $SO(3)$ symmetry. However,
a sphere cannot be tessellated into a finite number of identical cells
without breaking continuous rotational symmetry,
creating a fundamental obstacle for such numerical schemes.

The fuzzy-sphere regularization,
recently introduced for studying CFTs
in $(2+1)$-dimensional spacetime~\cite{Zhu_Han_Huffman_Hofmann_He_2023},
solves this problem.
The idea is to replace the algebra of functions on the sphere
with a finite-dimensional matrix algebra~\cite{JMadore1992}. 
This is
achieved by placing charged particles in the lowest Landau level on a two sphere pierced by a Dirac magnetic
monopole flux~\cite{Haldane_1983_FQHE,WU1976365}, which naturally gives a
finite-dimensional basis that respects rotational symmetry. 
This approach has been highly successful for extracting CFT data, as
demonstrated by its application to the Ising transition
in $(2+1)$-dimensional spacetime
\cite{Zhu_Han_Huffman_Hofmann_He_2023,Fuzzy_OPE,Fuzzy_4pt_correlators,Ising_CPT},
where it enabled precise determination of scaling dimensions, operator
product expansion coefficients, and correlation functions.
The fuzzy sphere has also been applied to study line defects,
impurities, and boundary CFTs~\cite{Fuzzy_Defects,Fuzzy_Impurities,Dedushenko2024},
as well as surface CFTs~\cite{Fuzzy_Surface_CFTs}, the
$g$-function~\cite{Fuzzy_g_function}, the entropic
$F$-function~\cite{Fuzzy_Entropic_F_function}, and deconfined quantum
critical points~\cite{Fuzzy_SO5,Yang2025Jul}.
Further work has explored Wilson--Fisher bilayers, explicit generators,
anyonic regularization, $\mathrm{Sp}(N)$ and Potts models,
noncommutative-circle physics, Yang-Lee type physics, free scalars and
tricriticality, Chern-Simons matter, crosscap coefficients, Majorana
bilayers, gauged Majorana and Ising theories, quantum rotor models and
$O(N)$ free-scalar and Wilson-Fisher theories on the fuzzy sphere, and
identifying the Ising conformal field theory from the ground-state energy
on the fuzzy
sphere~\cite{Conformal_Fuzzy_Wilson_Fisher_Content,Fan2024,Zhou2024Oct,Voinea2025Anyonic,Yang2025Jan,Han2025,Fan2025,ArguelloCruz2025,EliasMiro2025,FuzzyRealScalarHe,FuzzyRealScalarTaylor,Zhou2025Jul,Dong2025,Voinea2025,Zhou2025Sep,Wiese2025,Dey_O3_Rotor_2025,guo2025onfreescalarwilsonfisherconformal}.
The key advantage is that it allows
numerical studies of finite-size systems while preserving
the continuous symmetries needed for the state-operator
correspondence, giving direct access to the operator spectrum of the
CFT.

The remainder of this paper is organized as follows. In
Sec.~\ref{sec:model_methods}, we introduce the quantum
spin-1 XY model on the fuzzy sphere and describe its symmetries,
and explain how to determine
the critical point using conformal perturbation theory. In
Sec.~\ref{sec:results}, we present our results for the spectrum of
scaling dimensions and OPE coefficients at criticality, and discuss
the connection between Goldstone mode physics in the ordered phase and
the large charge expansion predictions for operators with large $U(1)$
charge. We conclude in Sec.~\ref{sec:conclusion}. Technical details
concerning Haldane pseudopotentials, conformal perturbation theory,
exact diagonalization, and DMRG are relegated to the appendices. The
Supplementary Material~\cite{SuppMat} contains complete tables of all
scaling dimensions and OPE coefficients obtained from our ED and DMRG
calculations.

Prior to our work, the Wilson-Fisher $O(2)$ CFT was studied using other numerical techniques. Notably, Refs.~\cite{Kos_Poland_Simmons-Duffin_Vichi_2016,O2Bootstrap_Chester,Liu_Simmons_Duffin_O2_spectrum,Reehorst:2019pzi, chester2025bootstrappingsimplestdeconfinedquantum} determined scaling dimensions of many primary operators and their OPE coefficients using the numerical conformal bootstrap \cite{Poland_2019,Poland:2022qrs,Rychkov:2023wsd}. Accurate Monte-Carlo results for scaling dimensions of several operators are also available \cite{Hasenbusch_2019_improvedclockmodel,Hasenbusch:2025yrl,2011PhRvB..84l5136H,Hasenbusch:2025qpz,hasenbusch2025precisionestimateslargecharge}. We will be comparing our results to these prior results. In some cases, we will be using the very precise conformal bootstrap results as inputs for our analysis. We also mention the epsilon-expansion and large-N results reviewed in \cite{HENRIKSSON20231}, although we do not rely on them here.

\begin{remark}
  \label{remark:previous_fuzzy_sphere_work_on_O2_CFT}
Previous work on the $O(2)$ Wilson-Fisher CFT on the fuzzy sphere includes Ref.~\cite{guo2025onfreescalarwilsonfisherconformal}, which studies the fuzzy-sphere $O(N)$ Wilson-Fisher CFT for $N=2,3,4$ in a model closely related to ours. The authors build a general realization of the $O(N)$ Wilson-Fisher CFT from the singlet and vector of $O(N)$, in the same spirit as the $O(3)$ Wilson-Fisher construction in Ref.~\cite{Dey_O3_Rotor_2025}, where some of the present authors also gave a general fuzzy-sphere model for the $O(N)$ Wilson-Fisher CFT. However, in this current work, we begin from a spin model on a square lattice and fuzzify it while preserving the spin symmetry. We regard this as a more direct construction of the $O(2)$ Wilson-Fisher CFT on the fuzzy sphere.

Reference~\cite{guo2025onfreescalarwilsonfisherconformal} gives evidence for an emergent $O(2)$ Wilson-Fisher CFT on the fuzzy sphere, but considers only systems with up to $11$ fermions and does not report the low-lying primary operator content. We provide that content here, including scaling dimensions and OPE coefficients. Aspects of this work were presented at~\cite{Lauchli_FuzzySphere_Talk_2023,Lauchli_FuzzySphere_Talk_2025,Lauchli_FuzzySphere_MeetsBootstrap_2025} before Ref.~\cite{guo2025onfreescalarwilsonfisherconformal} appeared.

\end{remark}

\section{Model and Methods}
\label{sec:model_methods}

\subsection{\texorpdfstring{$S=1$}{S=1} XY Model with \texorpdfstring{$O(2)$}{O(2)} Criticality}
\label{sec:S1_model}

The ferromagnetic quantum $S=1$ XY model with single-ion
anisotropy~\cite{spinonexy1} provides a concrete realization of a
quantum phase transition in the $(2+1)$-dimensional $O(2)$
Wilson-Fisher universality class. The Hamiltonian on a lattice
$\Lambda$
reads
\begin{subequations}
\label{eq:lattice hamiltonian}
\begin{equation} 
\widehat{H}^{(S=1)}_{XY}:=
-J
\sum_{\langle i,j\rangle}
\left(
\widehat{S}^{x}_{i}\,\widehat{S}^{x}_{j}
+
\widehat{S}^{y}_{i}\,\widehat{S}^{y}_{j}
+
\alpha\,
\widehat{S}^{z}_{i}\,\widehat{S}^{z}_{j}
\right)
+
D
\sum_{i\in\Lambda}
(\widehat{S}^{z}_{i})^{2},
\label{eq:lattice hamiltonian a}
\end{equation}
where the triplet of Hermitian operators
\begin{equation}
\widehat{S}^{a}_{i}=\left(\widehat{S}^{a}_{i}\right)^{\dagger},
\qquad
a=x,y,z,
\label{eq:lattice hamiltonian b}
\end{equation}
at the site $i\in\Lambda$
are the three quantum spin-1 operators and
$\langle i,j\rangle$ are directed nearest-neighbor bonds on the lattice
$\Lambda$.
These operators satisfy the $\mathfrak{su}(2)$ algebra
\begin{equation}
[\widehat{S}^{a}_{i}, \widehat{S}^{b}_{j}] =
\mathrm{i} \epsilon^{abc}\, \widehat{S}^{c}\,\delta^{\,}_{ij},
\qquad
i,j\in\Lambda,
\label{eq:lattice hamiltonian c}
\end{equation}
where $\epsilon^{abc}$ is the Levi-Civita symbol,
with the quadratic Casimir operator
\begin{equation}
\widehat{\bm{S}}^{2}_{i}=S(S+1)=2,
\qquad
i\in\Lambda.
\label{eq:lattice hamiltonian d}
\end{equation}
If we choose the quantization axis along the $z$-direction
of the quantum spin-1 space, they have the
quantum spin-1 matrix representations
\begin{align}
\widehat{S}^{x} =
\frac{1}{\sqrt{2}}
\begin{pmatrix}
 0 &+1 & 0\\
+1 & 0 &+1 \\
 0 &+1 & 0\\
\end{pmatrix},
\qquad
\widehat{S}^{y} =
\frac{1}{\sqrt{2}}
\begin{pmatrix}
0 & -\mathrm{i} & 0\\
+\mathrm{i} & 0 & -\mathrm{i} \\
0 & +\mathrm{i} & 0\\
\end{pmatrix},
\qquad
\widehat{S}^{z} =
\begin{pmatrix}
+1 & 0 & 0\\
0 & 0 & 0 \\
0 & 0 & -1\\
\end{pmatrix},
\label{eq:lattice hamiltonian e}
\end{align}
\end{subequations}
when their action is restricted to a single site.
The energy scale in Hamiltonian (\ref{eq:lattice hamiltonian a})
is set by the ferromagnetic exchange interaction with
coupling strength $J>0$ (the negative sign favors
ferromagnetic ordering). The dimensionless
parameter $-1<\alpha<1$
controls the strength of the Ising-like $S^{z}$ coupling
relative to the XY interactions. Finally, the energy scale $D>0$
is the single-ion anisotropy parameter that energetically favors the
$|S^{z}=0\rangle$ state at each site, creating an easy-plane anisotropy.

The Hamiltonian (\ref{eq:lattice hamiltonian a})
can be obtained from two different microscopic starting points.
First, it arises from truncating a Bose-Hubbard
model~\cite{fisher_PRB_89,hen_PRB_09,hen_PRB_10} to the subspace with
zero, one, or two bosons per site. Second, it can be derived by
truncating an $O(2)$ quantum rotor model to its two lowest angular
momentum sectors, analogous to the construction used for the $O(3)$
case in Ref.~\cite{Dey_O3_Rotor_2025}.

Reference~\cite{Whitsitt_Schuler_Henry_Lauchli_Sachdev_2017}
investigated the energy spectrum of the $O(2)$-symmetric $\phi^{4}$
field theory at the Wilson-Fisher fixed point on a torus, employing the
$\epsilon = 3 - d$ expansion, where $d$ denotes the number of spatial dimensions. They also performed exact diagonalization studies of the quantum spin-1 XY lattice Hamiltonian~(\ref{eq:lattice hamiltonian a})
on the square lattice with periodic boundary conditions.
The phase diagram as a function of $D/J$ was shown to
exhibit three distinct
regimes~\cite{spinonexy1}, which we describe here for the case
$\alpha=0$. For $D=0$, the absence of single-ion anisotropy allows
ferromagnetic long-range order in the $xy$ quantum spin-1 plane,
with spins spontaneously breaking the $SO(2)$ symmetry. At large values
of $D\gg J$, the single-ion anisotropy dominates, forcing each
quantum spin-1 into its eigenstate $|0\rangle_{i}$
with eigenvalue $0$ of $\widehat{S}^{z}_{i}$,
resulting in a gapped quantum paramagnetic product state
$\prod_{i\in\Lambda} |0\rangle_{i}$
with no long-range order. Between these two phases, at an
intermediate critical value $D=D_c$, the system undergoes a continuous
quantum phase transition belonging to the $(2+1)$-dimensional $O(2)$
Wilson-Fisher universality class.
As values of $|\alpha|<1$
do not affect the universality class of the transition,
we set $\alpha=0$
without loss of generality.

The symmetry group $G$ that leaves both Hamiltonian
(\ref{eq:lattice hamiltonian a})
and the $\mathfrak{su}(2)$ algebra
(\ref{eq:lattice hamiltonian c})
invariant is
\begin{subequations}
\label{eq:symmetry group quantum spin-1 Hamiltonian}
\begin{equation}
G=
G^{\,}_{\Lambda}\times
O(2)
\times\mathbb{Z}^{\mathrm{T}}_{2}.
\label{eq:symmetry group quantum spin-1 Hamiltonian a}
\end{equation}
Here,
$G^{\,}_{\Lambda}$ is the space group of the lattice $\Lambda$.
The internal symmetry group
\begin{equation}
O(2)\cong
SO(2)\rtimes\mathbb{Z}^{\mathrm{R}^{\,}_{xz}}_{2}
\label{eq:symmetry group quantum spin-1 Hamiltonian b}
\end{equation}
is a semidirect product.
The continuous symmetry group $SO(2)$ consists of
all global spin rotations
\begin{equation}
\widehat{U}^{z}(\theta):=
e^{-\mathrm{i}\theta\widehat{S}^{z}},
\qquad
0\leq\theta<2\pi,
\qquad
\widehat{S}^{z}:=
\sum_{i\in\Lambda}
\widehat{S}^{z}_{i},
\label{eq:symmetry group quantum spin-1 Hamiltonian c}
\end{equation}  
by $\theta$ about the $z$ axis in quantum spin-1 space,
i.e., conjugation of the quantum spin-1 operators by
$\widehat{U}^{z}(\theta)$ amounts to 
\begin{equation}
\widehat{S}^{x}_{i}
\mapsto
\cos\theta\,\widehat{S}^{x}_{i}+\sin\theta\,\widehat{S}^{y}_{i},
\qquad
\widehat{S}^{y}_{i}
\mapsto
-\sin\theta\,\widehat{S}^{x}_{i}+\cos\theta\,\widehat{S}^{y}_{i},
\qquad
\widehat{S}^{z}_{i}
\mapsto
\widehat{S}^{z}_{i},
\qquad
\forall i\in\Lambda.
\label{eq:symmetry group quantum spin-1 Hamiltonian d}
\end{equation}  
The finite symmetry group $\mathbb{Z}^{\mathrm{R}^{\,}_{xz}}_{2}$
is generated by the discrete rotation
\begin{equation}
\widehat{U}^{\,}_{\mathrm{R}^{\,}_{xz}}:=
e^{-\mathrm{i}\pi\widehat{S}^{y}},
\qquad
\widehat{S}^{y}:=
\sum_{i\in\Lambda}
\widehat{S}^{y}_{i},
\label{eq:symmetry group quantum spin-1 Hamiltonian e}
\end{equation}
by $\pi$ about the $y$ axis in quantum spin-1 space, i.e.,
conjugation of the quantum spin-1 operators by
$\widehat{U}^{\,}_{\mathrm{R}^{\,}_{xz}}$
amounts to 
\begin{equation}
\widehat{S}^{x}_{i}\mapsto
-\widehat{S}^{x}_{i},
\qquad
\widehat{S}^{y}_{i}\mapsto
+\widehat{S}^{y}_{i},
\qquad
\widehat{S}^{z}_{i}\mapsto
-\widehat{S}^{z}_{i},
\qquad
\forall i\in\Lambda.
\label{eq:symmetry group quantum spin-1 Hamiltonian f}
\end{equation}

The product (\ref{eq:symmetry group quantum spin-1 Hamiltonian b})
is semidirect because
$\widehat{U}^{\,}_{\mathrm{R}^{\,}_{xz}}$
conjugates
$\widehat{S}^{z}$ to
$-\widehat{S}^{z}$,
thereby inverting any $SO(2)$ rotation.
The group
$\mathbb{Z}^{\mathrm{T}}_{2}$ is generated by reversal of time that is
implemented by the antiunitary operator
\begin{equation}
\widehat{U}^{T}=e^{-\mathrm{i}\pi\widehat{S}^{y}}\,\mathsf{K},    
\qquad
\widehat{S}^{y}:=
\sum_{i\in\Lambda}
\widehat{S}^{y}_{i},
\label{eq:symmetry group quantum spin-1 Hamiltonian g}
\end{equation}
where $\mathsf{K}$ denotes complex conjugation of $\mathbb{C}$ numbers,
i.e., conjugation of the quantum spin-1 operators by
$\widehat{U}^{T}$ amounts to 
\begin{equation}
\widehat{S}^{x}_{i}\mapsto
-\widehat{S}^{x}_{i},
\qquad
  \forall  i\in\Lambda.
\label{eq:symmetry group quantum spin-1 Hamiltonian h}
\end{equation}
\end{subequations}
Reversal of time
commutes with both
$G^{\,}_{\Lambda}$ and $O(2)$
as it is antiunitary and
both lattice momentum and spin angular momentum are odd under reversal of time.

\subsection{Fuzzy-Sphere Implementation}
\label{sec:fuzzy implementation}

We now implement the quantum $S=1$ XY model on the fuzzy
sphere~\cite{Zhu_Han_Huffman_Hofmann_He_2023} to leverage
the state-operator correspondence for extracting conformal data.
This is done by
using charged fermions with internal quantum spin $S=1$ (i.e.,
each fermion carries the flavor $\sigma = 0, \pm 1$)
living on a fuzzy-sphere geometry with a $4\pi s$ magnetic monopole at the
center. Each fermion also carries a label for its
orbital momentum $m = -s, -s+1, \cdots,s-1,s$
stemming from the $2s+1$-fold degeneracy of each
lowest Landau level (LLL) orbital. The fermionic creation
$\hat{c}^{\dagger}_{m,\sigma}$
and annihilation
$\hat{c}^{\,}_{m',\sigma'}$
operators thus
satisfy the canonical anticommutation
relations 
\begin{subequations}
\label{eq:fermion_algebra}
\begin{equation}
\{\hat{c}^{\,}_{m,\sigma}, \hat{c}^{\dagger}_{m',\sigma'}\} =
\delta^{\,}_{m,m'}\delta^{\,}_{\sigma,\sigma'},
\quad
\{\hat{c}^{\,}_{m,\sigma}, \hat{c}^{\,}_{m',\sigma'}\} = 0,
\quad
\{\hat{c}^{\dagger}_{m,\sigma}, \hat{c}^{\dagger}_{m',\sigma'}\} = 0,
\label{eq:fermion_algebra a}
\end{equation}
and we use the compact notation
\begin{equation}
\hat{\bm{c}}^{\dagger}_{m} =
(\hat{c}^{\dagger}_{m,1}, \hat{c}^{\dagger}_{m,0}, \hat{c}^{\dagger}_{m,-1})
\label{eq:fermion_algebra b}
\end{equation}
for the spinor of creation operators
with orbital momentum
$m = -s, -s+1, \cdots,+s-1,+s$ along the quantization axis.
The single-particle state
\begin{equation}
|m,\sigma\rangle\equiv\hat{c}^{\dagger}_{m,\sigma}\,|0\rangle,
\qquad
\hat{c}^{\,}_{m,\sigma}\,|0\rangle=0,
\qquad
m=-s,\cdots,+s,
\qquad
\sigma=-1,0,+1,
\label{eq:fermion_algebra c}
\end{equation}
\end{subequations}
where $|0\rangle$ is the state annihilated by all fermionic
annihilation operators,
describes a point particle with quantum spin-1
constrained to move on a sphere of radius
$R\propto\sqrt{2s+1}$
with (1) the orbital Casimir eigenvalue $s(s+1)$ and the orbital momentum $m$
along the quantization axis in orbital space
and (2) the spin-1 Casimir eigenvalue $1(1+1)$
and the spin $\sigma$ along the quantization axis in spin-1 space.

Generalizing the construction of the transverse field Ising model
in $(2+1)$-dimensional spacetime
in Ref.~\cite{Zhu_Han_Huffman_Hofmann_He_2023},   we consider the following many-body fermionic
Hamiltonian for the fuzzy sphere of radius $R\propto\sqrt{2s+1}$:
\begin{subequations}
\label{eq:H_fuzzy_sphere}
\begin{align}
    \label{eq:H_total}
    &
    \widehat{H}(R,D) :=
    \widehat{H}^{\,}_{00}
    +
    \widehat{H}^{\,}_{xy}
    +
    \widehat{H}^{\,}_{D},
    \\[6pt]
    \label{eq:H_00}
    &
    \widehat{H}^{\,}_{00} :=
    \sum_{m_1,m_2,m=-s}^{s} V^{\,}_{m_1,m_2,m_2-m,m_1+m}
    (\cdagbf{m_1}\,\cbf{m_1+m})(\cdagbf{m_2}\,\cbf{m_2-m}), \\[6pt]
    \label{eq:H_xy}
    &
    \begin{aligned}
    \widehat{H}^{\,}_{xy} := 
    - \frac{1}{2}\sum_{m_1,m_2,m=-s}^{s} V^{\,}_{m_1,m_2,m_2-m,m_1+m}
    \Big[&(\cdagbf{m_1}\,S^{x}\,\cbf{m_1+m})(\cdagbf{m_2}\,S^{x}\,\cbf{m_2-m}) \\
    &+ (\cdagbf{m_1}\,S^{y}\,\cbf{m_1+m})(\cdagbf{m_2}\,S^{y}\,\cbf{m_2-m})\Big],
    \end{aligned}\\[6pt]
    \label{eq:H_D}
    &
    \widehat{H}^{\,}_{D} :=
    D \sum_{m=-s}^{s} \cdagbf{m}\,(S^{z})^{2}\,\cbf{m},
\end{align}
where $\bm{S}\equiv(S^{x},S^{y},S^{z})^{\mathsf{T}}$
is the pseudovector of spin-1 matrices defined in
Eq.~\eqref{eq:lattice hamiltonian e}.
The couplings $V^{\,}_{m_1,m_2,m_3,m_4}\in\mathbb{R}$ and $D\geq0$ carry
units of energy and are both assumed to be much smaller in magnitude than the
single-particle energy gap separating the LLL from all other Landau levels.
Throughout this work the fermionic charge is conserved: we place $N=2s+1$
fermions, one per orbital on average, into the $3(2s+1)$ single-particle states
$|m,\sigma\rangle$ of Eq.~\eqref{eq:fermion_algebra c}, so that the filling
fraction is held fixed at
\begin{equation}
  \nu:=\frac{N}{3\,(2s+1)}=\frac{2s+1}{3\,(2s+1)}=\frac{1}{3}.
  \label{eq:nu}
\end{equation}
\end{subequations}
The Hamiltonian $\widehat{H}(R,D)$ thus governs the quantum dynamics of
$N$ fermions constrained to move on the surface of a sphere of radius
$R\propto\sqrt{2s+1}$ in the presence of a $4\pi s$ magnetic monopole. Each of
the three terms $\widehat{H}^{\,}_{00}$, $\widehat{H}^{\,}_{xy}$, and
$\widehat{H}^{\,}_{D}$ in $\widehat{H}(R,D)$ has a distinct physical role and
preserves specific symmetries, as we now explain.

Both the density-density interaction
$\widehat{H}^{\,}_{00}$ and the XY interaction $\widehat{H}^{\,}_{xy}$
depend on the couplings 
$V^{\,}_{m_1,m_2,m_3,m_4}$.
We chose these couplings to be the projections
to the LLL of the repulsive interaction
\begin{equation}
V^{\,}_{0}\,
\delta^{(2)}(\bm{n}-\bm{n}')
+
V^{\,}_{1}\,
\nabla^{2}\delta^{(2)}(\bm{n}-\bm{n}'),
\qquad
V^{\,}_{0},V^{\,}_{1}\,>0,
\end{equation}
between two fermions whose positions are labeled by the unit vectors
$\bm{n}$ and $\bm{n}'$ on the unit sphere $S^{2}$
(following the construction in Ref.\
\cite{Zhu_Han_Huffman_Hofmann_He_2023} 
and Appendix\
\ref{app:pseudopotentials}).
Here, we have only retained 
Haldane's pseudopotentials $V^{\,}_{l}$
for a pair of fermions with relative angular momentum $l=0,1$
on the sphere.

The density-density interaction
$\widehat{H}^{\,}_{00}$ in Eq.~\eqref{eq:H_00} is a spin-independent
two-body repulsion that preserves both spatial $SO(3)$ rotational
symmetry and internal $O(2)$ spin symmetry. It penalizes
fermions with the same spin $\sigma=-1,0,+1$ from being too close to each other
on the sphere of radius $R\propto\sqrt{2s+1}$.

The XY interaction $\widehat{H}^{\,}_{xy}$ in Eq.~\eqref{eq:H_xy} implements
the ferromagnetic exchange coupling in the $xy$ quantum spin-1 plane
while preserving both spatial $SO(3)$ and internal $O(2)$ symmetries.
It favors spin alignment and drives the system toward the ordered phase
with spontaneously broken internal $O(2)$ symmetry in the limit of
small $D$ relative to $V^{\,}_{0}$ and $V^{\,}_{1}$.

The single-ion anisotropy $\widehat{H}^{\,}_{D}$ in
Eq.~\eqref{eq:H_D} is a one-body term that preserves both spatial 
$SO(3)$ and internal $O(2)$ symmetries. It penalizes the $|\sigma=\pm1\rangle$
states at each orbital, favoring $|\sigma=0\rangle$, thereby driving
the system toward the quantum paramagnetic phase in the limit of large
$D$ relative to $V^{\,}_{0}$ and $V^{\,}_{1}$.

The symmetry group $G$ that leaves both Hamiltonian
(\ref{eq:H_fuzzy_sphere})
and the fermion algebra
(\ref{eq:fermion_algebra a})
invariant is
\begin{subequations}
\label{eq:symmetry group quantum fuzzy Hamiltonian}
\begin{equation}
G=
SO(3)
\times
O(2)
\times
\left[
U(1)
\rtimes
\mathbb{Z}^{\mathrm{T}}_{2}
\right].
\label{eq:symmetry group quantum fuzzy Hamiltonian a}
\end{equation}
The orbital continuous symmetry group $SO(3)$
(the pseudopotentials $V^{\,}_{l}$ are $SO(3)$-invariant by construction)
is generated by 
\begin{equation}
\widehat{U}(\bm{\alpha}):=
e^{-\mathrm{i}\bm{\alpha}\cdot\widehat{\bm{L}}},
\qquad
\bm{\alpha}\in\mathbb{R}^{3},
\qquad
\widehat{\bm{L}}:=
\sum_{\sigma,\sigma'=0,\pm1}\
\sum^{+s}_{m,m'=-s}
\hat{c}^{\dag}_{m,\sigma}\,
\bm{L}^{\,}_{m,m'}\,
\delta^{\,}_{\sigma,\sigma'}\,
\hat{c}^{\,}_{m',\sigma'},
\label{eq:symmetry group quantum fuzzy Hamiltonian b}
\end{equation}
where the three $(2s+1)\times(2s+1)$ matrices defining $\bm{L}$ have the 
matrix elements 
\begin{equation}
L^{\pm}_{m',m}\equiv
\left(L^{x}\pm\mathrm{i}L^{y}\right)^{\,}_{m',m}=
\delta^{\,}_{m',m\pm1}\,\sqrt{(s\mp m)(s\pm m+1)},
\qquad
L^{z}_{m',m}=m\,\delta^{\,}_{m',m}.
\label{eq:symmetry group quantum fuzzy Hamiltonian c}
\end{equation}
The internal symmetry group
\begin{equation}
O(2)=SO(2)\rtimes\mathbb{Z}^{\mathrm{R}^{\,}_{xz}}_{2}
\label{eq:symmetry group quantum fuzzy Hamiltonian d}
\end{equation}
is the semidirect product of the global continuous symmetry group
$SO(2)$
with the infinitesimal generator
\begin{equation}
\widehat{S}^{z}:=
\sum^{+s}_{m=-s}\
\sum_{\sigma,\sigma'=0,\pm1}\,
\hat{c}^{\dag}_{m,\sigma}\,
S^{z}_{\sigma,\sigma'}\,
\hat{c}^{\,}_{m,\sigma'}
\label{eq:symmetry group quantum fuzzy Hamiltonian e}  
\end{equation}
and the discrete symmetry group 
$\mathbb{Z}^{\mathrm{R}^{\,}_{xz}}_{2}$
that is generated by the rotation
\begin{equation}
\widehat{U}^{\,}_{R^{\,}_{xz}}:=
e^{-\mathrm{i}\pi\widehat{S}^{y}},
\qquad
\widehat{S}^{y}:=
\sum^{+s}_{m =-s}\
\sum_{\sigma,\sigma'=0,\pm1}\,
\hat{c}^{\dag}_{m,\sigma}\,
S^{y}_{\sigma,\sigma'}\,
\hat{c}^{\,}_{m,\sigma'}.
\label{eq:symmetry group quantum fuzzy Hamiltonian f}  
\end{equation}
by $\pi$ about the $y$ axis in quantum spin-1 space
and is explicitly given by
\begin{equation}
\widehat{U}^{\,}_{R^{\,}_{xz}}=
\sum^{+s}_{m=-s}\ 
\sum_{\sigma,\sigma'=0,\pm1}
\hat{c}^{\dag}_{m,\sigma}\,
(-1)^{1-\sigma}\,\delta^{\,}_{\sigma',-\sigma}\,
\hat{c}^{\,}_{m,\sigma'}.
\label{eq:symmetry group quantum fuzzy Hamiltonian g}  
\end{equation} 
The group $U(1)$ is generated
\begin{equation}
\widehat{U}^{\,}_{U(1)}(\alpha):=
e^{\mathrm{i}\alpha\widehat{N}},
\qquad
\alpha\in[0,2\pi[,
\label{eq:symmetry group quantum fuzzy Hamiltonian h}  
\end{equation}
with the fermion-number operator    
\begin{equation}      
\widehat{N}:=
\sum^{+s}_{m=-s}\
\sum_{\sigma,\sigma'=0,\pm1}
\hat{c}^{\dag}_{m,\sigma}\,
\delta^{\,}_{\sigma,\sigma'}\,  
\hat{c}^{\,}_{m,\sigma'}
\label{eq:symmetry group quantum fuzzy Hamiltonian i}  
\end{equation}
as the infinitesimal generator.
The group
$\mathbb{Z}^{\mathrm{T}}_{2}$ is generated by reversal of time
whose representation is
\begin{equation}
\widehat{U}^{\,}_{T}:= 
e^{+\mathrm{i}\pi\left(\widehat{L}^{y}+\widehat{S}^{y}\right)}\,
\mathsf{K},
\label{eq:symmetry group quantum fuzzy Hamiltonian j}
\end{equation}
\end{subequations}
where $\mathsf{K}$ denotes complex conjugation of $\mathbb{C}$ numbers as reversal of time must be implemented with an antiunitary operator.
Hence, it commutes with the symmetries generated by
the orbital angular momentum $\widehat{\bm{L}}$ or by $\widehat{S}^{z}$.
Since the fermion-number operator is even under conjugation by
$\widehat{U}^{\,}_{T}$, reversal of time takes
$\widehat{U}^{\,}_{U(1)}(\alpha)$ to
$\widehat{U}^{\,}_{U(1)}(-\alpha)$, which is why it enters as the semidirect product with the $U(1)$ symmetry group generated by the total fermion number operator.  

In the fuzzy sphere regularized Ising model studied in Ref.~\cite{Zhu_Han_Huffman_Hofmann_He_2023}, the system is at half-filling ($\nu = 1/2$), which leads to a particle-hole symmetry. In the infrared limit, this particle-hole symmetry becomes the spacetime parity symmetry of the CFT. In contrast, our system is at one-third filling ($\nu = 1/3$), so there is no particle-hole symmetry. The spacetime parity symmetry of the $O(2)$ Wilson-Fisher CFT still emerges in the infrared. However, because our model does not have a microscopic symmetry which would directly correspond to the spacetime parity, we cannot easily recover the spacetime parity symmetry eigenvalue of the CFT operators.

Energy eigenstates are labeled by the quantum numbers
$L$, whereby $\widehat{\bm{L}}^{2}=L(L+1)$;
$S^{z}$ (eigenvalue of $\widehat{S}^{z}$);
and, in the $S^{z}=0$ sector,
the parity $P=\pm1$ (eigenvalue of
$\widehat{U}^{\,}_{\mathrm{R}^{\,}_{xz}}$). For $S^z\ne 0$ we have double degeneracy, with states $\pm S^z$ interchanged by $\widehat{U}^{\,}_{\mathrm{R}^{\,}_{xz}}$.
The fermion number $N=2s+1$ is fixed by the filling fraction
and does not serve as a label.
The thermodynamic limit is
$R\propto\sqrt{N}\uparrow\infty$.

In this work, we restrict to $l \leq 1$ in the pseudopotential
expansion, such that only $V^{\,}_{0}$ and $V^{\,}_{1}$ are active. The parameter
$D$ serves as our tuning parameter. At small values of $D$,
the system is in the ordered phase,
while at large values of $D$, it is in the paramagnetic phase,
with a quantum critical point at an intermediate value $D = D_c$.

\subsection{Critical Point Determination}

We first assume the existence of a quantum critical point
at $D=D_{\mathrm{c}}$
at which the Hamiltonian $\widehat{H}(R,D)$ defined by Eq.\
(\ref{eq:H_fuzzy_sphere})
exhibits the conformal invariance of the
(2+1)D $O(2)$ Wilson-Fisher universality class
in the thermodynamic limit $R\propto\sqrt{N}\uparrow\infty$.
If we denote eigenstates of $\widehat{H}(R,D)$
by the label $\mathrm{o}$ (see Table \ref{tab:primaries_table}),
the state-operator
correspondence~\cite{Rychkov_2017}
states
that the eigenenergy $E_{\mathrm{o}}(R,D_c)$ of $\widehat{H}(R,D_c)$
measured relative to that of the ground-state energy $E_{gs}(R,D_c)$
scales like
\begin{align}
\delta E_{\mathrm{o}}(R,D_c)\equiv
E_{\mathrm{o}}(R,D_c)-E_{gs}(R,D_c)\sim
\frac{c}{R}\,
\Delta_{\mathrm{o}}
\qquad
(\hbar=1)
\label{eq:deltaE_o at criticality}
\end{align}
in the thermodynamic limit $R\propto\sqrt{N}\uparrow\infty$.
Here, the radius $R$ of the fuzzy sphere is interpreted as
the radius of compactification in the radial quantization
of a CFT, $\Delta_{\mathrm{o}}$ is the scaling dimension
of the CFT operator labeled by $\mathrm{o}$,
and the characteristic microscopic speed $c$
becomes the speed of light in the CFT.

If $\lim_{R\propto\sqrt{N}\uparrow\infty}\widehat{H}(R,D_c)$
at quantum criticality is perturbed by
the perturbation with the small coupling
$\delta D=D-D_{\mathrm{c}}\neq0$, then
the energy shift to first-order in perturbation theory
is captured by adding the single relevant scalar operator
$\varepsilon$ (see Table \ref{tab:primaries_table})
with the coupling $g_{\varepsilon}(R, D)$,
together with irrelevant operators
to the (2+1)D $O(2)$ Wilson-Fisher critical point.

When $R$ is finite and
motivated by first-order conformal perturbation theory (CPT)%
~\cite{Icosahedron,Ising_CPT},
we make the scaling ansatz 
\begin{subequations}
\label{eq:deltaE_o close to criticality}
\begin{align}
\delta E_{\mathrm{o}}(R,D)=
\frac{c(R,D)}{R}\,
\Delta_{\mathrm{o}}(R)
+ 4 \pi
 g_{\varepsilon}(R, D)\,
f_{\mathrm{o}\varepsilon\mathrm{o}}(R).
\label{eq:deltaE_o close to criticality a}
\end{align}
Here,
\begin{equation}
\Delta_{\mathrm{o}}(R)=
\Delta_{\mathrm{o}}
+
\mathcal{O}(R^{-\omega})
\label{eq:deltaE_o close to criticality b}
\end{equation}
and
\begin{equation}
f_{\mathrm{o}\varepsilon\mathrm{o}}(R)=
f_{\mathrm{o}\varepsilon\mathrm{o}}
+
\mathcal{O}(R^{-\omega})
\label{eq:deltaE_o close to criticality c}
\end{equation}
\end{subequations}
deviate from their respective quantum-critical values
$\Delta_{\mathrm{o}}$
and
$f_{\mathrm{o}\varepsilon\mathrm{o}}$
after the thermodynamic limit $R\propto\sqrt{N}\uparrow\infty$ has been taken due to the presence of the leading irrelevant operator perturbing the CFT
with the scaling exponent
$\omega>0$. 

When slightly detuning away from the critical point, the coefficient $g_{\varepsilon}(R, D)$ is expected to increase with the volume according to $\sim R^{2-\Delta_\varepsilon}$ \cite{Icosahedron,Ising_CPT}. This RG running effect was checked in \cite{Ising_CPT}. But in this work we will not use this theoretically known dependence of $g_{\varepsilon}(R, D)$ on the volume.

The interpretation of 
$f_{\mathrm{o}\varepsilon\mathrm{o}}$
defined by taking the thermodynamic limit
$R\propto\sqrt{N}\uparrow\infty$ 
in Eq.\ (\ref{eq:deltaE_o close to criticality c})
is the following.
(i)
It is the (universal) OPE coefficient corresponding
to fusing operators
$\mathrm{o}$ and $\varepsilon$ into $\mathrm{o}$,
if $\mathrm{o}$ labels a scalar primary field.
(ii)
It is proportional to
$f_{\mathrm{o}_{\mathrm{p}}\varepsilon\mathrm{o}_{\mathrm{p}}}(R)$
with the proportionality constant a known function of 
$\Delta_{\mathrm{o}_{\mathrm{p}}}$ and $\Delta_{\varepsilon}$,
if $\mathrm{o}$ labels the descendant of a primary field labeled by
$\mathrm{o}_{\mathrm{p}}$%
~\cite{Icosahedron,Ising_CPT}.
(iii) Finally,
it is a combination of OPE coefficients \cite{Ising_CPT}
if $\mathrm{o}$ labels a field with non-vanishing conformal spin.

Our strategy to establish the existence of
the (2+1)D $O(2)$ Wilson-Fisher critical point
from numerical diagonalization of $\widehat{H}(R,D)$
proceeds in two steps. First, we choose two
eigenstates (CFT operators) of $\widehat{H}(R,D)$ labeled by
$\mathrm{o}_{1}\equiv\sigma$
and
$\mathrm{o}_{2}\equiv\partial_{\mu}\sigma$,
where $\sigma$ labels the most relevant primary operator
(quantum numbers $S^{z}=1$, $L=0$),
and $\partial_{\mu}\sigma$ labels its first descendant
(quantum numbers $S^{z}=1$, $L=1$).
This choice is motivated by the empirical
observation that these levels are typically least affected by
finite-size corrections and irrelevant operators~\cite{Ising_CPT}. For
these operators, we insert into Eq.~(\ref{eq:deltaE_o close to
  criticality}) the central values of the conformal bootstrap (CB) determinations:
\begin{equation}
\Delta_{\sigma}(R)=\Delta_{\sigma}=0.519088,\quad f_{\sigma\varepsilon\sigma}(R)= f_{\sigma\varepsilon\sigma}=0.687126
\end{equation}
from Table~\ref{tab:primaries_table} and
Table~\ref{tab:opes_with_primaries}, respectively.\footnote{In this project we chose to use CB determinations as reference values, although for scaling dimensions MC determinations are in many cases somewhat more precise, see Table~\ref{tab:primaries_table}.} In other words, we use these as reference values, neglecting $\mathcal{O}(R^{-\omega})$ corrections to scaling for these quantities.
We similarly neglect corrections to scaling for $\Delta_{\partial_{\mu}\sigma}=1+\Delta_{\sigma}=1.519088$ together
with $f_{\partial_{\mu}\sigma,\varepsilon,\partial_{\mu}\sigma}$,
which is related to $f_{\sigma\varepsilon\sigma}$ through the
descendant relation~\cite{Icosahedron,Ising_CPT}. We thus get a system of two
linear equations in the two unknowns $c(R,D)$ and $g_{\varepsilon}(R, D)$, which is solved
for each value of $R,D$ to determine the speed of light and the coupling
as a function of $R,D$.  

Next, for each $R=\sqrt{N}$, we determine the root
$D_{\mathrm{c}}(R)$ to $g_{\varepsilon}(R, D)=0$. The values $D_{\mathrm{c}}(R)$
obtained from this procedure for different system sizes $N$
are summarized in Table~\ref{tab:critical_points}.

Further, for any eigenstate (CFT operator) of $\widehat{H}(R,D)$ labeled by $\mathrm{o}$, we
proceed in two steps using the ansatz Eq.~(\ref{eq:deltaE_o close to criticality}).
First, we set $D=D_{\mathrm{c}}(R)$, i.e.\ the value of $D$ at which $g_{\varepsilon}(R,D)=0$.
At this value the perturbation term drops out of Eq.~(\ref{eq:deltaE_o close to criticality a}),
so that $\delta E_{\mathrm{o}} = c(R,D_{\mathrm{c}})\,\Delta_{\mathrm{o}}(R)/R$;
using the already-determined $c(R,D_{\mathrm{c}})$ we then read off $\Delta_{\mathrm{o}}(R)$.
Second, we vary $D$ away from $D_{\mathrm{c}}(R)$ while keeping $R$ fixed.
This turns on a nonzero $g_{\varepsilon}(R,D)$, and the shift in $\delta E_{\mathrm{o}}(R,D)$
relative to the $g_{\varepsilon}=0$ value is given by $g_{\varepsilon}(R,D)\,f_{\mathrm{o}\varepsilon\mathrm{o}}(R)$,
from which we extract $f_{\mathrm{o}\varepsilon\mathrm{o}}(R)$.

We do not extrapolate with a fitted $\mathcal{O}(R^{-\omega})$ ansatz; $R^{-\omega}$ is only used as the plotting abscissa in the figures, with $\omega\equiv\Delta_{\varepsilon'}-3$ fixed from the bootstrap value of $\Delta_{\varepsilon'}$. Tabulated scaling dimensions and OPE coefficients are taken at the largest available $R$.

\begin{remark}
  \label{remark:choice_of_sigma}
We use $\sigma$ and $\partial_{\mu}\sigma$ to determine the critical point and the speed of light $c$. Among the low-lying primaries, these are the least affected by the leading irrelevant operator $\varepsilon'$, because the OPE coefficient $f_{\sigma\varepsilon'\sigma}$ is very small compared to how $\varepsilon'$ perturbs the other low-lying operators:
  \begin{align}
     f_{\sigma \varepsilon' \sigma} = 0.0393(3), \quad
    f_{\varepsilon \varepsilon' \varepsilon} = 1.27(1), \quad
    f_{t \varepsilon' t} = 0.600(4)\,.\footnotemark
    \label{eq:irrelevant_ope_coefficients_with_epsilon_prime}
  \end{align}
By contrast, the relevant perturbation from $\varepsilon$ affects $\sigma$ and $\partial_{\mu}\sigma$ substantially, as Table~\ref{tab:opes_with_primaries} shows. They are therefore natural choices for determining the critical point and $c$: they respond strongly to departures from criticality, yet remain comparatively insensitive to irrelevant perturbations.
\end{remark}
 \footnotetext{We are grateful to David Poland for communicating to us preliminary unpublished values of these OPE coefficients extracted from the data of \cite{Liu_Simmons_Duffin_O2_spectrum}. }

\section{Results}
\label{sec:results}

\subsection{Scaling Dimensions}
\label{sec:scaling_dimensions}

\begin{figure}[htbp]
    \centering
    \includegraphics[width=0.935\textwidth]{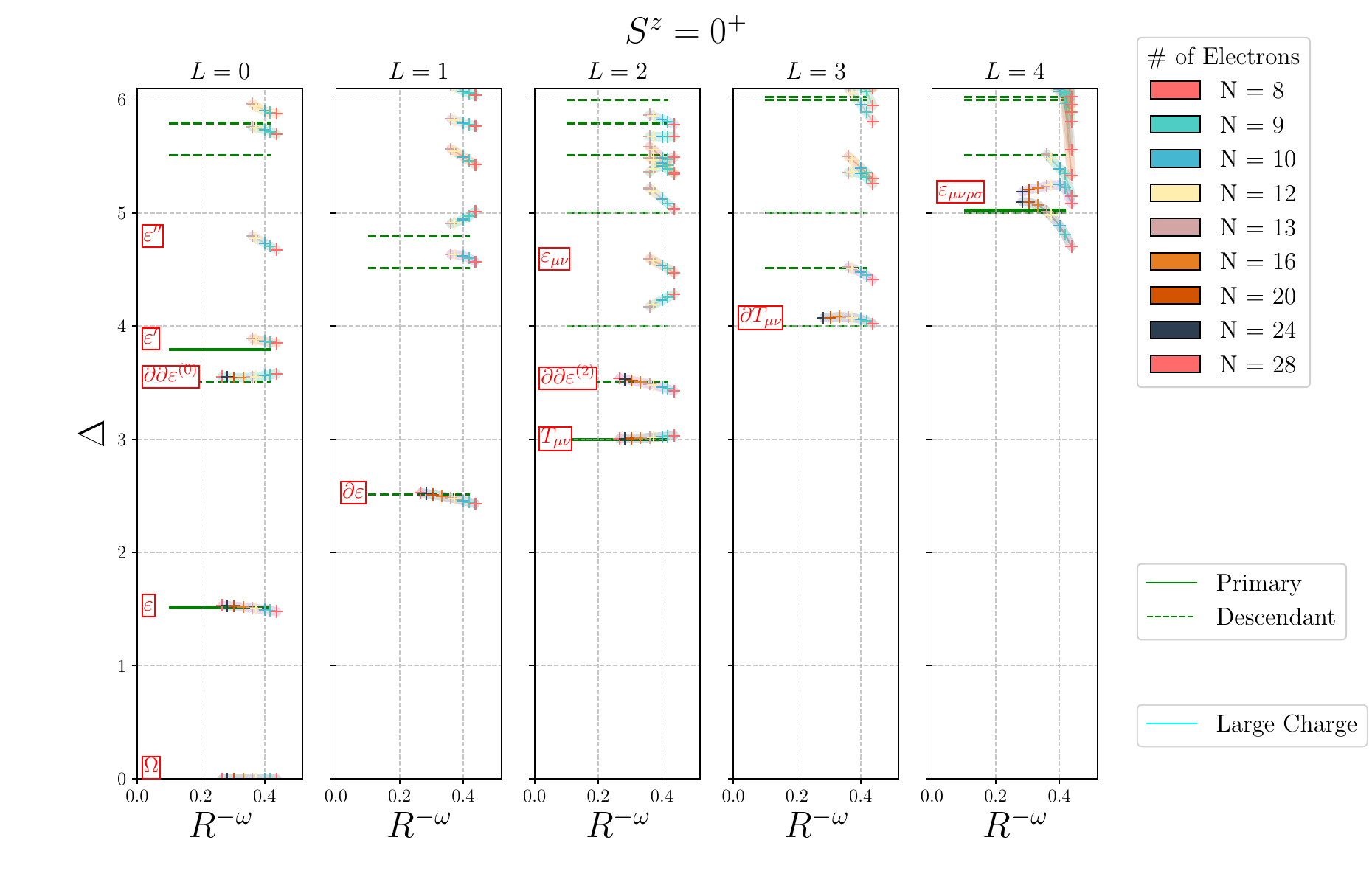}
    \includegraphics[width=0.935\textwidth]{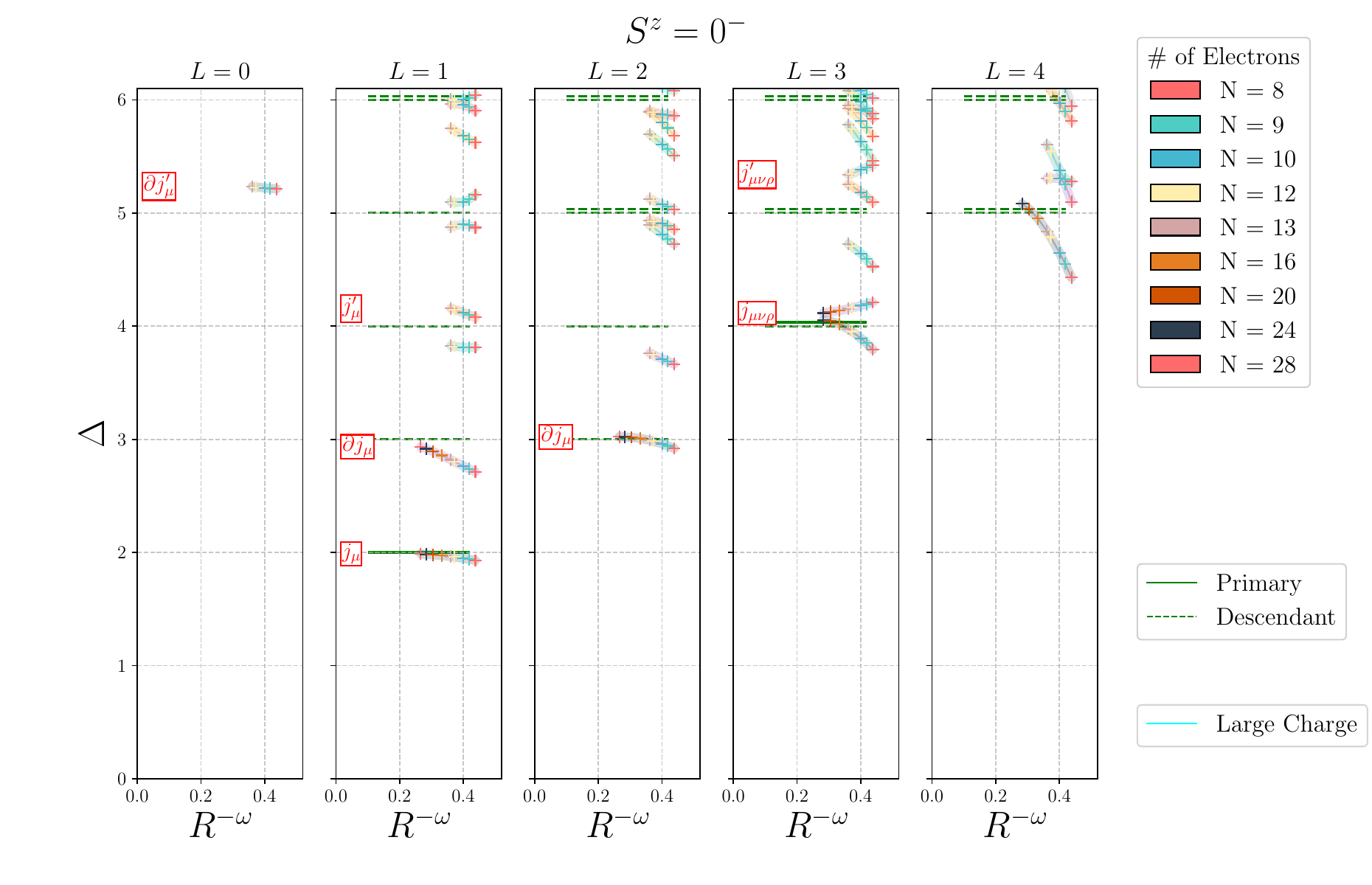}
    \caption{
      ED and DMRG spectrum for the $O(2)$ quantum
      spin-1 model (\ref{eq:H_fuzzy_sphere})
      in $(2+1)$-dimensional spacetime
      at $V^{\,}_{0} = 4.0, V^{\,}_{1}=1.0$ and $D$ at
      criticality.
      Dependence on $R^{-\omega}$ of the dimensionless number
      $\Delta_{\mathrm{o}}(R)$ for eigenstates with eigenvalue
      $S^{z}=0$ for the projection
      (\ref{eq:symmetry group quantum fuzzy Hamiltonian e})
      of the total spin along the quantization axis
      resolved by the eigenvalues $\pm$
      of the parity operator
      (\ref{eq:symmetry group quantum fuzzy Hamiltonian g})
      is shown.
      The limiting value
      $\Delta_{\mathrm{o}}$ as $R\propto\sqrt{N}\uparrow\infty$
      is interpreted as the
      scaling dimension of the CFT operator labeled by $\mathrm{o}$.
      Green lines indicate CB predictions (or exact values)
      when available.
      The choice of eigenstates is organized by
      the label $L$ of the Casimir operator for the total
      orbital angular momentum
      (\ref{eq:symmetry group quantum fuzzy Hamiltonian b}).
\label{fig:O2SpectrumSz0}}
\end{figure}

\begin{figure}[htbp]
    \centering
    \includegraphics[width=0.935\textwidth]{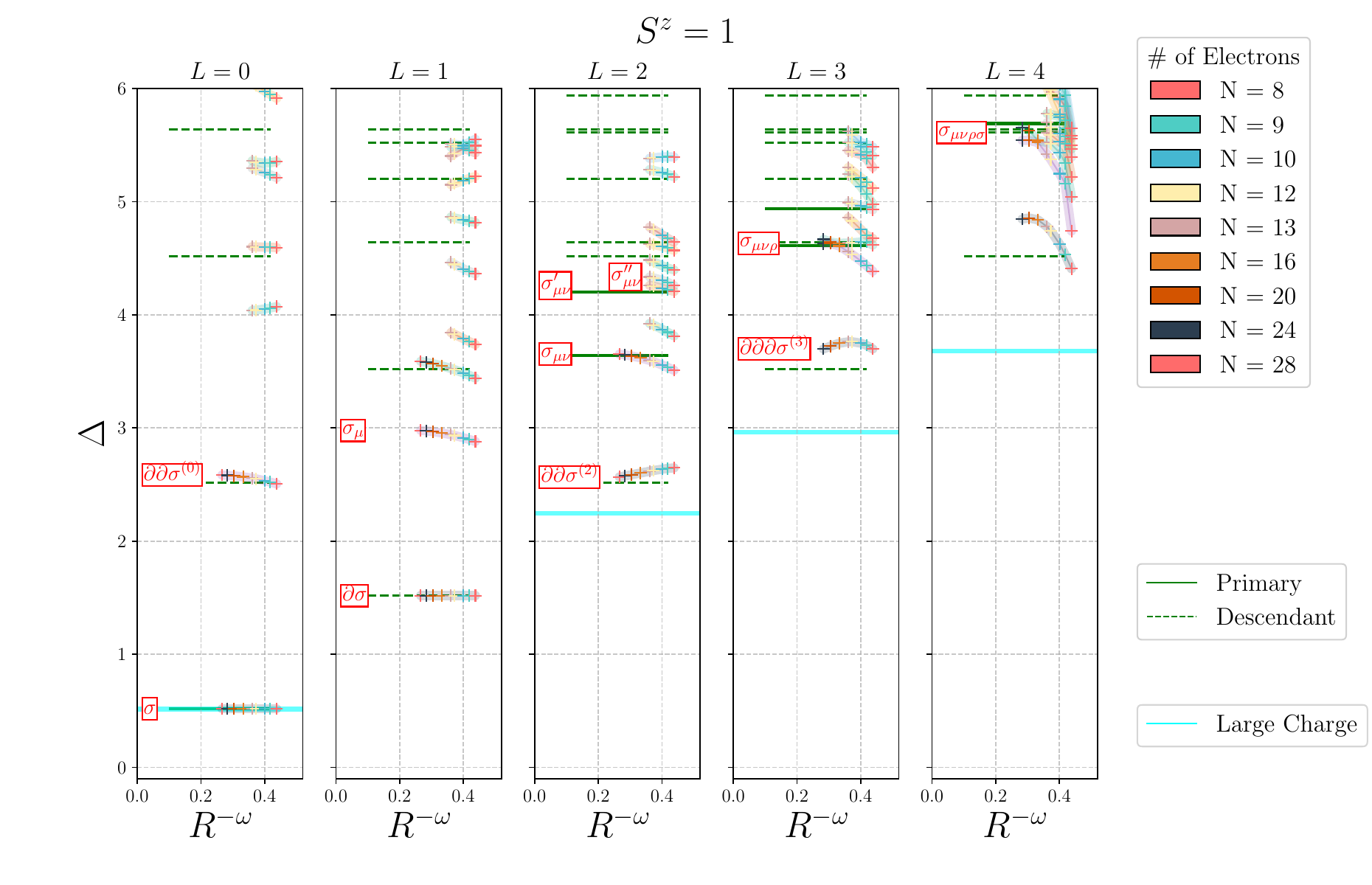}
    \includegraphics[width=0.935\textwidth]{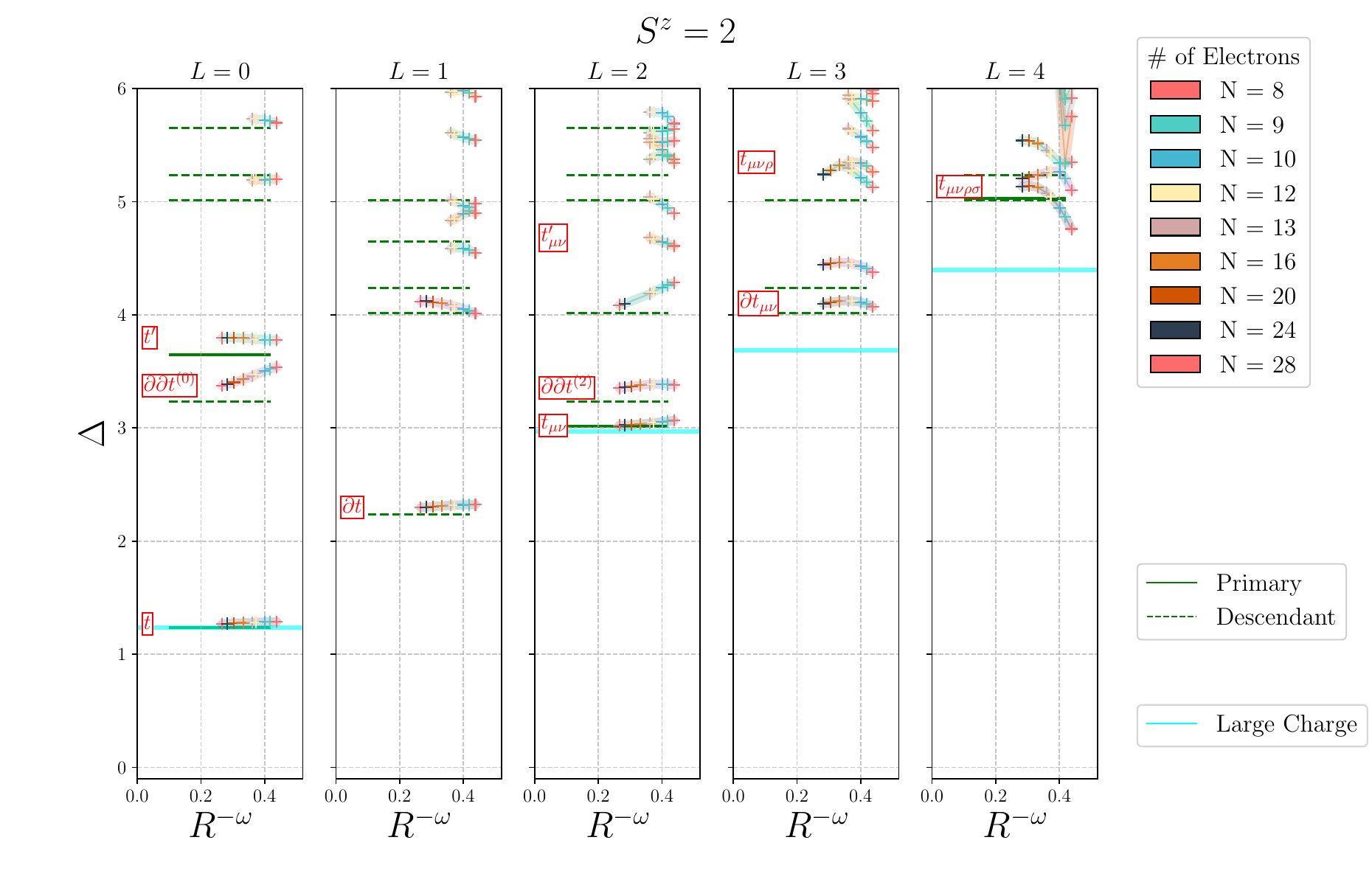}
    \caption{
      ED and DMRG spectrum for the $O(2)$ quantum
      spin-1 model (\ref{eq:H_fuzzy_sphere})
      in $(2+1)$-dimensional spacetime
      at $V^{\,}_{0} = 4.0, V^{\,}_{1}=1.0$ and $D$ at
      criticality, showing the $S^{z}=1$ and $S^{z}=2$ sectors
      for the projection
      (\ref{eq:symmetry group quantum fuzzy Hamiltonian e})
      of the total spin along the quantization axis. The
      cyan lines are primary operators predicted by the large charge
      expansion~\cite{Hellerman15,Monin17,Cuomo2020}. The lowest cyan
      line in each charge sector corresponds to the ground state
      prediction from the large charge expansion, while higher cyan
      lines represent phonon primaries. For $S^{z}=2$,
      the first (lowest) primary is used to fix the Wilson
      coefficients in the large charge expansion, so no cyan
      prediction line appears for these states.
\label{fig:O2SpectrumSz12}}
\end{figure}

\begin{figure}[htbp]
    \centering
    \includegraphics[width=0.935\textwidth]{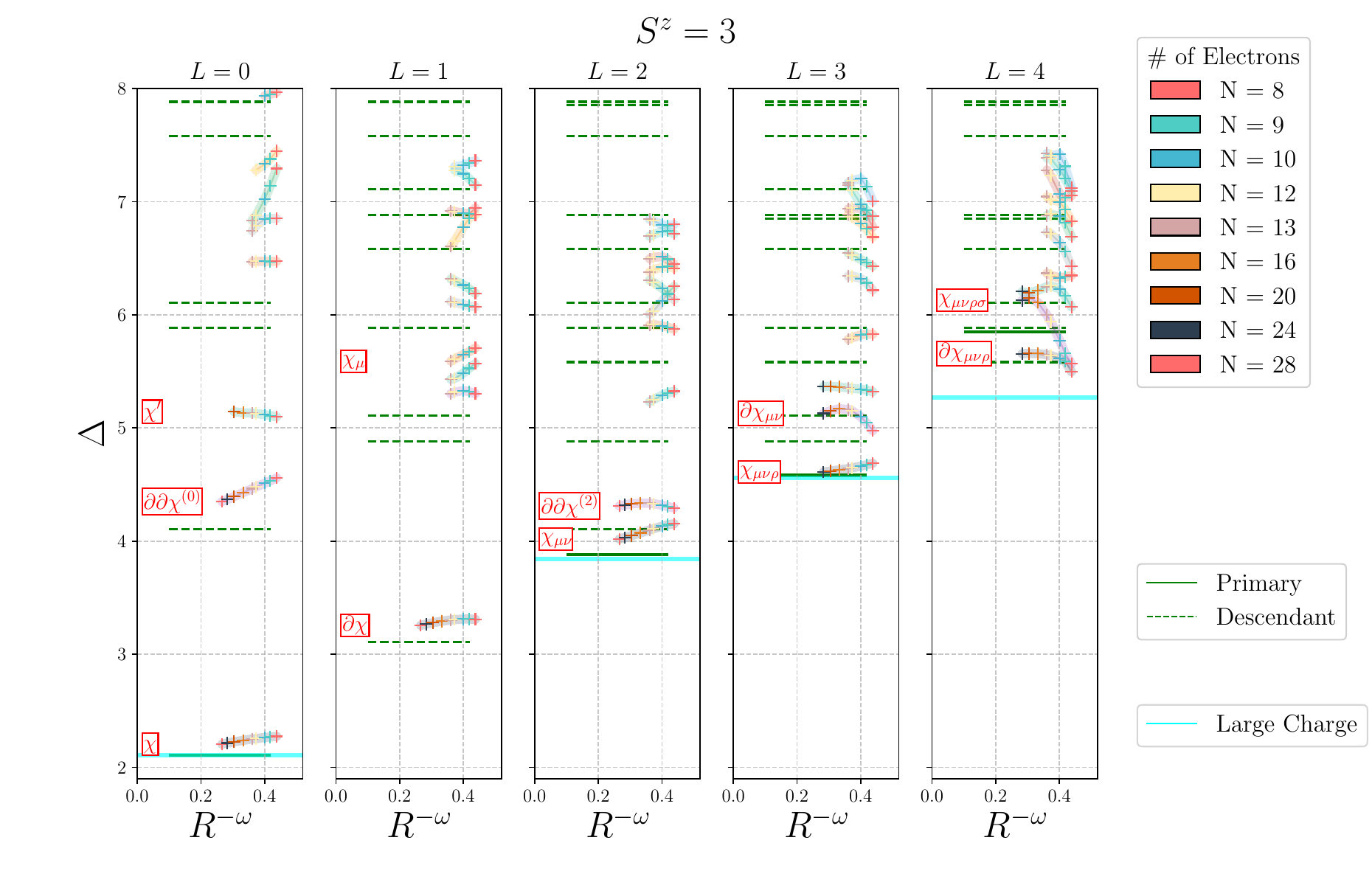}
    \includegraphics[width=0.935\textwidth]{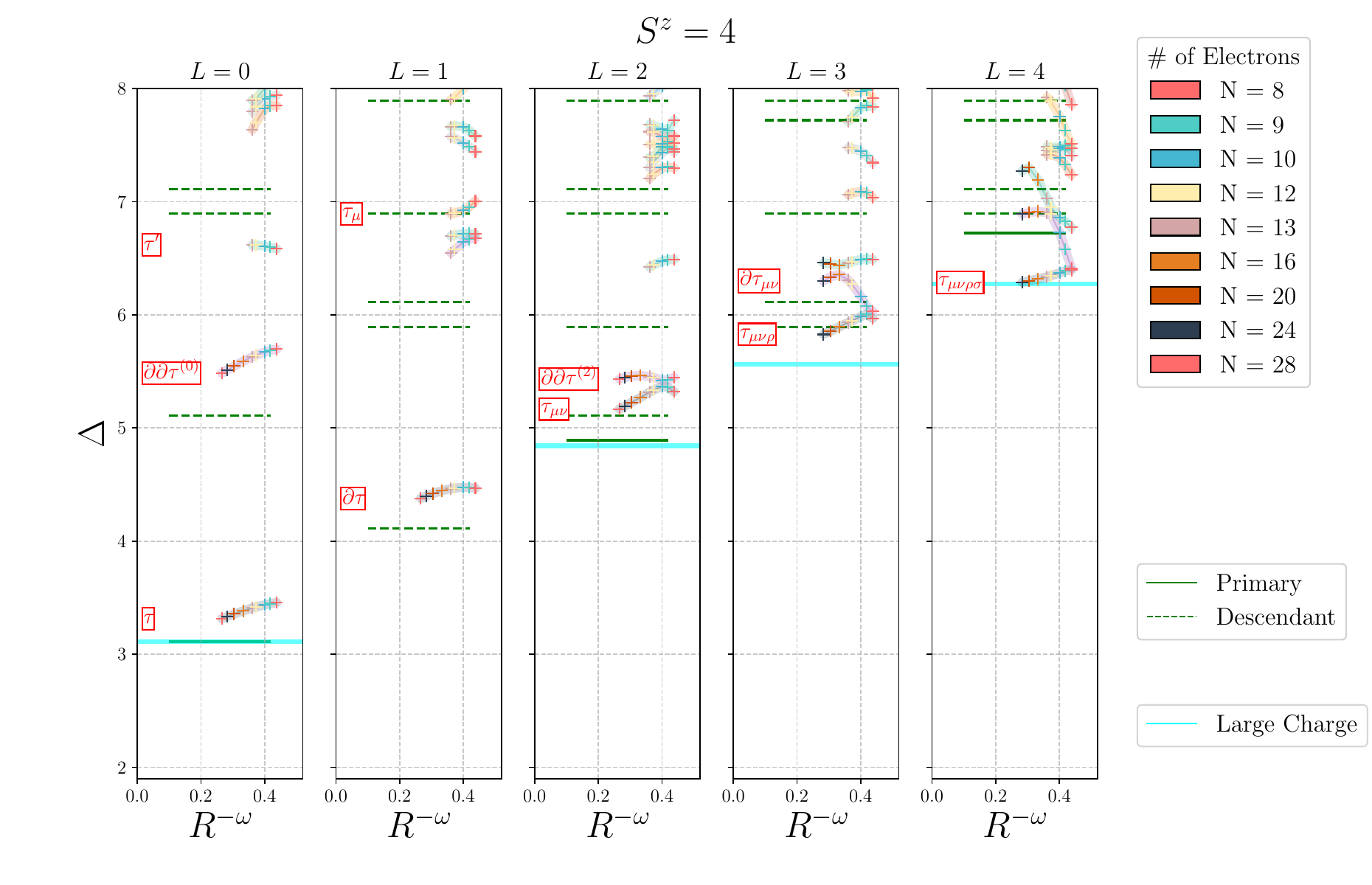}
    \caption{
      ED and DMRG spectrum for the $O(2)$ quantum
      spin-1 model (\ref{eq:H_fuzzy_sphere})
      in $(2+1)$-dimensional spacetime
      at $V^{\,}_{0} = 4.0, V^{\,}_{1}=1.0$ and $D$ at
      criticality, showing the $S^{z}=3$ and $S^{z}=4$ sectors
      for the projection
      (\ref{eq:symmetry group quantum fuzzy Hamiltonian e})
      of the total spin along the quantization axis.
      The cyan lines are primary operators predicted by the large charge
      expansion. As in Fig.~\ref{fig:O2SpectrumSz12}, the lowest cyan
      line represents the ground state prediction, and higher lines
      are phonon primaries. For $S^{z}=4$, the first primary is used
      to fix Wilson coefficients, hence no cyan prediction line
      appears for it.
\label{fig:O2SpectrumSz34}}
\end{figure}

The dependence on $R^{-\omega}$ of the dimensionless number
$\Delta_{\mathrm{o}}(R)$ defined by the scaling ansatz
(\ref{eq:deltaE_o close to criticality}) with $D$ chosen to be the
root of $g_{\varepsilon}(R, D)=0$ for each value of $R$ is shown in
Figs.~\ref{fig:O2SpectrumSz0}, \ref{fig:O2SpectrumSz12}, and
\ref{fig:O2SpectrumSz34} for a selection of eigenstates of
$\widehat{H}(R)$ obtained from ED and DMRG\footnote{For the larger system sizes, the coupling $g_{\varepsilon}$ at the estimated critical point was not always exactly zero, typically it was of order $10^{-3}$, because the scan over $D$ used a relatively coarse grid. In these cases, we evaluated $\Delta_{\mathrm{o}}(R, D)$ at several nearby values of $D$ and interpolated to find $\Delta_{\mathrm{o}}(R)$ at $g_{\varepsilon}(R, D) = 0$. Since operator eigenvalues were already computed at various $D$ values for the OPE coefficient analysis, this interpolation did not incur any additional computational cost.
}(for details about the ED and DMRG implementation see Appendix~\ref{app:ED} and Appendix~\ref{app:DMRG}).
Here, $\Delta_{\varepsilon'}$ is the scaling dimension of the leading
irrelevant perturbation, and $\omega\equiv\Delta_{\varepsilon'}-3$ is
its deviation from marginality. Our estimates for the scaling
dimensions $\Delta_{\mathrm{o}}$, obtained from the largest available
ED and DMRG system sizes, are reported in
Table~\ref{tab:primaries_table} for all identified primary operators.

{\tiny
\renewcommand{\arraystretch}{0.88}
\begin{longtable}{|ccc|c|c|c|c|c|c|}
  \caption{
    Primary operators of the $O(2)$ Wilson-Fisher CFT, as determined from the quantum spin-1 model
    (\ref{eq:H_fuzzy_sphere}) in $(2+1)$-dimensional spacetime
    at $V^{\,}_{0} = 4.0, V^{\,}_{1}=1.0$ and $D$ at criticality.
    The quantum number
    $S^{z}$ is the eigenvalue of the projection
    (\ref{eq:symmetry group quantum fuzzy Hamiltonian e})
    along the quantization axis of the total spin; it maps to the $U(1)$ global charge of the CFT.
    The quantum number $L$ labels the Casimir operator for the
    total orbital angular momentum
    (\ref{eq:symmetry group quantum fuzzy Hamiltonian b}); it is the spin of the CFT primary.
    The quantum number
    $\pm$
    labels the eigenvalue of the parity operator
    (\ref{eq:symmetry group quantum fuzzy Hamiltonian g})
    in the sector of the Hilbert space with $S^{z}=0$.
    The index $I$ corresponds to the position of the energy eigenstate within each symmetry sector; e.g.~$I=2,4$ for $\varepsilon,\varepsilon'$ in the $L=0$ column of Fig.~\ref{fig:O2SpectrumSz0}, where $I=1,3$ are the unit operator and the $\partial^2 \varepsilon$ descendant.
    Scaling dimensions are extracted via ED and DMRG, the subscript indicating the largest system size (fermion number) where we have the data.
    These are compared, where possible, to conformal bootstrap (CB)  \cite{O2Bootstrap_Chester,Liu_Simmons_Duffin_O2_spectrum,chester2025bootstrappingsimplestdeconfinedquantum} and Monte Carlo (MC)  \cite{Hasenbusch_2019_improvedclockmodel,Hasenbusch:2025yrl,2011PhRvB..84l5136H,Hasenbusch:2025qpz,hasenbusch2025precisionestimateslargecharge}.
    The primary shown in red is the one whose CB central value is used to determine the critical point and the speed of light $c$.
    For $S^{z}=0^{+},0^{-},1,2,3,4,5,6,7$, we denote the primary operators by $\varepsilon,j,\sigma,t,\chi,\tau,\rho,\lambda,\eta$, respectively.
    Subscripts $\mu, \nu, \rho, \sigma, \tau$ indicate tensor indices: no subscript for a scalar ($L=0$), one for a vector ($L=1$), two for a rank-2 tensor ($L=2$), etc. If multiple primaries exist for a given $S^{z}$ and $L$, a prime $'$ marks the second lowest, with additional primes for successively higher primaries. 
  The values in the parentheses for CB and MC results indicate the errors. The CB errors in boldface are rigorous; those with asterisk ($^*$) are nonrigorous since obtained via the extremal functional method \cite{El-Showk:2012vjm} from a collection of allowed points.
  The status ``confirmed'' means that the same operator has previously been identified by CB, MC, or an exact conservation law; ``new'' means that no comparable previous determination is available. Both statuses refer to primary assignments in the present spectrum.
  }\label{tab:primaries_table}\\ \hline
      $S^{z}$ & $L$ & $I$ &$\mathrm{o}$ & $\Delta$(ED) & $\Delta$(DMRG) &
    $\Delta$(CB) & $\Delta$(MC) & Status
    \\
    \hline
    \hline
    $0^{+}$ & 0 & 2 & $\mathbf{\varepsilon}$ & 1.510065$_{13}$ & 1.532935$_{28}$ & 1.51136\textbf{(22)} \cite{O2Bootstrap_Chester} & 1.51128(5) \cite{Hasenbusch:2025yrl} & confirmed \\
    \hline
    $0^{+}$ & 0 & 4 & $\varepsilon'$ & 3.887999$_{13}$ & ~ & 3.794(8$^*$) \cite{Liu_Simmons_Duffin_O2_spectrum}
    & 3.789(4) \cite{Hasenbusch_2019_improvedclockmodel} & confirmed\\
    \hline
    $0^{+}$ & 0 & 5 & $\varepsilon''$ & 4.796534$_{13}$ & ~ & ~ & ~ & new\\
    \hline
    $0^{+}$ & 2 & 1 & $T_{\mu \nu}$ & 3.015277$_{13}$ & 3.005932$_{28}$ & 3 (exact) & ~ & confirmed\\
    \hline
    $0^{+}$ & 2 & 4 & $\varepsilon_{\mu\nu}$ & 4.596819$_{13}$ & ~ & ~ & ~ & new\\
    \hline
    $0^{+}$ & 4 & 2 & $\varepsilon_{\mu\nu\rho\sigma}$ & 5.237991$_{13}$ & 5.184636$_{24}$ & 5.0254(4$^*$) \cite{Liu_Simmons_Duffin_O2_spectrum} & ~ & confirmed \\
    \hline
    $0^{-}$ & 1 & 1 & $j_{\mu}$ & 1.964120$_{13}$ & 1.987539$_{28}$ & 2 (exact)& ~ & confirmed\\
    \hline
    $0^{-}$ & 1 & 4 & $j'_{\mu}$ & 4.156422$_{13}$ & ~ & ~ & ~ & new\\
    \hline
    $0^{-}$ & 3 & 2 & $j_{\mu\nu\rho}$ & 4.154722$_{13}$ & 4.115489$_{24}$ & 4.0343(4$^*$) \cite{Liu_Simmons_Duffin_O2_spectrum} & ~ & confirmed\\
    \hline
    $0^{-}$ & 3 & 5 & $j'_{\mu\nu\rho}$ & 5.337671$_{13}$ & ~ & ~ & ~ & new\\
    \hline
    1 & 0 & 1 & $\mathbf{\sigma}$ & \red{0.519088}$_{13}$ & \red{0.519088}$_{28}$ & \red{0.519088}\textbf{(22)} \cite{O2Bootstrap_Chester} & 0.51908(1)  \cite{Hasenbusch:2025yrl} & confirmed\\
    \hline
    1 & 1 & 2 & $\sigma_{\mu}$ & 2.941156$_{13}$ & 2.978990$_{28}$ &  2.950\cite{chester2025bootstrappingsimplestdeconfinedquantum} & ~ & confirmed\\
    \hline
    1 & 2 & 2 & $\sigma_{\mu\nu}$ & 3.597402$_{13}$ & 3.652814$_{28}$ & 3.64(3$^*$) \cite{Liu_Simmons_Duffin_O2_spectrum} & ~ & confirmed\\
    \hline
    1 & 2 & 4 & $\sigma'_{\mu\nu}$ & 4.259708$_{13}$ & ~ & 4.20(2$^*$) \cite{Liu_Simmons_Duffin_O2_spectrum} & ~ & confirmed\\
    \hline
    1 & 2 & 5 & $\sigma''_{\mu\nu}$ & 4.334637$_{13}$ & ~ & ~ & ~ & new\\
    \hline
    1 & 3 & 2 & $\sigma_{\mu\nu\rho}$ & 4.557210$_{13}$ & 4.629527$_{24}$ & 4.614(8$^*$) \cite{Liu_Simmons_Duffin_O2_spectrum} & ~ & confirmed\\
    \hline
    1 & 4 & 4 & $\sigma_{\mu\nu\rho\sigma}$ & 5.611690$_{13}$ & ~ & 5.690(8$^*$) \cite{Liu_Simmons_Duffin_O2_spectrum} & ~ & confirmed\\
    \hline
    2 & 0 & 1 & $t$ & 1.282029$_{13}$ & 1.267871$_{28}$ & 1.23629\textbf{(11)} \cite{O2Bootstrap_Chester} & 1.23630(12) \cite{hasenbusch2025precisionestimateslargecharge} & confirmed\\
    \hline
    2 & 0 & 3 & $t'$ & 3.789029$_{13}$ & 3.795163$_{28}$ & 3.650(2$^*$) \cite{Liu_Simmons_Duffin_O2_spectrum}& ~ & confirmed\\
    \hline
    2 & 2 & 1 & $t_{\mu\nu}$ & 3.042022$_{13}$ & 3.023389$_{28}$ & 3.01537(3$^*$) \cite{Liu_Simmons_Duffin_O2_spectrum} & ~ & confirmed\\
    \hline
    2 & 2 & 4 & $t'_{\mu\nu}$ & 4.680865$_{13}$ & ~ & ~ & ~ & new\\
    \hline
    2 & 3 & 4 & $t_{\mu\nu\rho}$ & 5.350336$_{13}$ & ~ & ~ & ~ & new\\
    \hline
    2 & 4 & 1 & $t_{\mu\nu\rho\sigma}$ & 5.075681$_{13}$ & 5.133003$_{24}$ & 5.0308(6$^*$) \cite{Liu_Simmons_Duffin_O2_spectrum} & ~ & confirmed\\
    \hline
    3 & 0 & 1 & $\chi$ & 2.251853$_{13}$ & 2.207272$_{28}$ & 2.1086(3$^*$) \cite{Liu_Simmons_Duffin_O2_spectrum}& 2.10833(23) \cite{hasenbusch2025precisionestimateslargecharge} & confirmed\\
    \hline
    3 & 0 & 3 & $\chi'$ & 5.131775$_{13}$ & 5.143698$_{20}$ & ~ & ~ & new\\
    \hline
    3 & 1 &  &  &  &  & 2.078 \cite{chester2025bootstrappingsimplestdeconfinedquantum} &  & \\
    \hline
    3 & 1 & 4 & $\chi_{\mu}$ & 5.590976$_{13}$ & ~ & ~ & ~ & new\\
    \hline
    3 & 2 & 1 & $\chi_{\mu\nu}$ & 4.099157$_{13}$ & 4.018932$_{28}$ & 3.883(25$^*$) \cite{Liu_Simmons_Duffin_O2_spectrum}& ~ & confirmed\\
    \hline
    3 & 3 & 1 & $\chi_{\mu\nu\rho}$ & 4.642049$_{13}$ & 4.611154$_{24}$ & 4.582(1$^*$) \cite{Liu_Simmons_Duffin_O2_spectrum} & ~ & confirmed\\
    \hline
    3 & 4 & 2 & $\chi_{\mu\nu\rho\sigma}$ & 5.996864$_{13}$ & 6.128203$_{24}$ & 5.851(7$^*$) \cite{Liu_Simmons_Duffin_O2_spectrum} & ~ & confirmed\\
    \hline
    3 & 5 & 2 & $\chi_{\mu\nu\rho\sigma\lambda}$ & 6.620238$_{13}$ & 6.782848$_{24}$ & ~ & ~ & new\\
    \hline
    4 & 0 & 1 & $\tau$ & 3.408996$_{13}$ & 3.316242$_{28}$ & 3.11535(73$^*$) \cite{Liu_Simmons_Duffin_O2_spectrum} & 3.11203(34) \cite{hasenbusch2025precisionestimateslargecharge} & confirmed\\
    \hline
    4 & 0 & 3 & $\tau'$ & 6.616791$_{13}$ & ~ & ~ & ~ & new\\
    \hline
    4 & 1 & 4 & $\tau_{\mu}$ & 6.894136$_{13}$ & ~ & ~ & ~ & new\\
    \hline
    4 & 2 & 1 & $\tau_{\mu\nu}$ & 5.316778$_{13}$ & 5.167872$_{28}$ & 4.893(5$^*$) \cite{Liu_Simmons_Duffin_O2_spectrum} & ~ & confirmed\\
    \hline
    4 & 3 & 1 & $\tau_{\mu\nu\rho}$ & 5.932580$_{13}$ & 5.826713$_{24}$ & ~ & ~ & new\\
    \hline
    4 & 4 & 1 & $\tau_{\mu\nu\rho\sigma}$ & 6.336543$_{13}$ & 6.287773$_{24}$ & 6.72(2$^*$) \cite{Liu_Simmons_Duffin_O2_spectrum} & ~ & confirmed\\
    \hline
    5 & 0 & 1 & $\rho$ & 4.740359$_{13}$ & ~ & ~ & 4.23195(47) \cite{hasenbusch2025precisionestimateslargecharge} & confirmed\\
    \hline
    5 & 0 & 3 & $\rho'$ & 8.236994$_{13}$ & ~ & ~ & ~ & new\\
    \hline
    5 & 2 & 1 & $\rho_{\mu\nu}$ & 6.684267$_{13}$ & ~ & ~& ~ & new\\
    \hline
    5 & 3 & 1 & $\rho_{\mu\nu\rho}$ & 7.367956$_{13}$ & ~ & ~ & ~ & new\\
    \hline
    5 & 4 & 1 & $\rho_{\mu\nu\rho\sigma}$ & 7.831272$_{13}$ & ~ & ~ & ~ & new\\
    \hline
    5 & 5 & 1 & $\rho_{\mu\nu\rho\sigma\lambda}$ & 8.136111$_{13}$ & ~ & ~ & ~ & new\\
    \hline
    6 & 0 & 1 & $\lambda$ & 6.236087$_{13}$ & ~ & ~ & 5.45714(62) \cite{hasenbusch2025precisionestimateslargecharge} & confirmed\\
    \hline
    6 & 2 & 1 & $\lambda_{\mu\nu}$ & 8.149670$_{13}$ & ~ & ~ & ~ & new\\
    \hline
    6 & 3 & 1 & $\lambda_{\mu\nu\rho}$ & 8.878337$_{13}$ & ~ & ~ & ~ & new\\
    \hline
    7 & 0 & 1 & $\eta$ & 7.888051$_{13}$ & ~ & ~ & 6.77934(78) \cite{hasenbusch2025precisionestimateslargecharge} & confirmed\\
    \hline
    7 & 2 & 1 & $\eta_{\mu\nu}$ & 9.716965$_{13}$ & ~ & ~ & ~ & new\\
    \hline
    7 & 3 & 1 & $\eta_{\mu\nu\rho}$ & 10.391125$_{13}$ & ~ & ~ & ~ & new\\
    \hline%
\end{longtable}%
}

{\tiny
\begin{longtable}{|ccc|c|c|c|c|}
  \caption{
    OPE coefficients $f_{\mathrm{o}\varepsilon\mathrm{o}}$ in the $O(2)$ Wilson-Fisher CFT, as extracted via ED and DMRG from the $O(2)$ quantum spin-1 model (\ref{eq:H_fuzzy_sphere})
    in $(2+1)$-dimensional spacetime
    at $V^{\,}_{0} = 4.0, V^{\,}_{1}=1.0$, by varying
    $D$ around its critical value.  Only the primary OPE coefficients are reported. The same conventions as in
    Table~\ref{tab:primaries_table} are used for labeling operators. The conformal bootstrap (CB) results from \cite{O2Bootstrap_Chester} are given for comparison. The CB central value of $f_{\sigma\varepsilon\sigma}$, shown in red, is used as a reference in ED and DMRG, to calibrate the extraction of other OPE coefficients. See Appendix \ref{sec:J} for the discussion of normalization of the OPE coefficient $f_{j_\mu\varepsilon j_\mu}$.}
  \label{tab:opes_with_primaries}\\ \hline
  $S^{z}$ & $L$ & $I$ &$\mathrm{o}$ & $f_{\mathrm{o}\varepsilon \mathrm{o}}$(ED) & $f_{\mathrm{o}\varepsilon \mathrm{o}}$(DMRG) & $f_{\mathrm{o}\varepsilon \mathrm{o}}$(CB) \\
\hline\hline
$0^{+}$ & 0 & 2 & $\varepsilon$ & 0.856146$_{12}$ & 0.818711$_{28}$ & 0.830914(32*)~\cite{O2Bootstrap_Chester} \\
\hline
$0^{+}$ & 0 & 4 & $\varepsilon'$ & 1.545649$_{12}$ & ~ & ~ \\
\hline
$0^{+}$ & 0 & 5 & $\varepsilon''$ & 1.634283$_{12}$ & ~ & ~ \\
\hline
$0^{+}$ & 2 & 1 & $T_{\mu\nu}$ & 0.573559$_{12}$ & 0.581426$_{28}$ & ~ \\
\hline
$0^{+}$ & 2 & 4 & $\varepsilon_{\mu\nu}$ & 1.708548$_{12}$ & ~ & ~ \\
\hline
$0^{+}$ & 4 & 2 & $\varepsilon_{\mu\nu\rho\sigma}$ & 0.118165$_{12}$ & 0.279247$_{24}$ & ~ \\
\hline
$0^{-}$ & 1 & 1 & $j_{\mu}$ & 0.976163$_{12}$ & 0.979486$_{28}$ & $\pm 0.9674(60)$~\cite{Reehorst:2019pzi} \\
\hline
$0^{-}$ & 1 & 4 & $j'_{\mu}$ & 1.744587$_{12}$ & ~ & ~ \\
\hline
$0^{-}$ & 3 & 2 & $j_{\mu\nu\rho}$ & 0.265343$_{12}$ & 0.547630$_{24}$ & ~ \\
\hline
$0^{-}$ & 3 & 5 & $j'_{\mu\nu\rho}$ & 0.451834$_{12}$ & ~ & ~ \\
\hline
$1$ & 0 & 1 & $\sigma$ & 0.687126$_{12}$ & 0.687126$_{28}$ & \red{0.687126}(27*)~\cite{O2Bootstrap_Chester} \\
\hline
$1$ & 1 & 2 & $\sigma_{\mu}$ & 1.389879$_{12}$ & 1.362427$_{28}$ & ~ \\
\hline
$1$ & 2 & 2 & $\sigma_{\mu\nu}$ & 1.264959$_{12}$ & 1.280060$_{28}$ & ~ \\
\hline
$1$ & 2 & 4 & $\sigma'_{\mu\nu}$ & 0.962586$_{12}$ & ~ & ~ \\
\hline
$1$ & 2 & 5 & $\sigma''_{\mu\nu}$ & 0.882706$_{12}$ & ~ & ~ \\
\hline
$1$ & 3 & 2 & $\sigma_{\mu\nu\rho}$ & 1.075526$_{12}$ & 0.989197$_{24}$ & ~ \\
\hline
$1$ & 4 & 4 & $\sigma_{\mu\nu\rho\sigma}$ & 0.796941$_{12}$ & ~ & ~ \\
\hline
$2$ & 0 & 1 & $t$ & 1.256118$_{12}$ & 1.255077$_{24}$ & 1.25213(14*)~\cite{O2Bootstrap_Chester} \\
\hline
$2$ & 0 & 3 & $t'$ & 1.593230$_{12}$ & 1.659525$_{24}$ & ~ \\
\hline
$2$ & 2 & 1 & $t_{\mu\nu}$ & 0.507020$_{12}$ & 0.533299$_{24}$ & ~ \\
\hline
$2$ & 2 & 4 & $t'_{\mu\nu}$ & 1.761551$_{12}$ & ~ & ~ \\
\hline
$2$ & 3 & 4 & $t_{\mu\nu\rho}$ & 0.128497$_{12}$ & ~ & ~ \\
\hline
$2$ & 4 & 1 & $t_{\mu\nu\rho\sigma}$ & 0.147018$_{12}$ & 0.222921$_{24}$ & ~ \\
\hline
$3$ & 0 & 1 & $\chi$ & 1.763969$_{12}$ & 1.770280$_{28}$ & ~ \\
\hline
$3$ & 0 & 3 & $\chi'$ & 2.193333$_{12}$ & 2.573009$_{20}$ & ~ \\
\hline
$3$ & 1 & 4 & $\chi_{\mu}$ & 1.075807$_{12}$ & ~ & ~ \\
\hline
$3$ & 2 & 1 & $\chi_{\mu\nu}$ & 1.101125$_{12}$ & 1.177578$_{28}$ & ~ \\
\hline
$3$ & 3 & 1 & $\chi_{\mu\nu\rho}$ & 0.868749$_{12}$ & 0.904905$_{24}$ & ~ \\
\hline
$3$ & 4 & 2 & $\chi_{\mu\nu\rho\sigma}$ & 0.724608$_{12}$ & 0.812521$_{24}$ & ~ \\
\hline
$3$ & 5 & 2 & $\chi_{\mu\nu\rho\sigma\lambda}$ & 0.572864$_{12}$ & 0.578391$_{24}$ & ~ \\
\hline
$4$ & 0 & 1 & $\tau$ & 2.229994$_{12}$ & 2.244990$_{28}$ & ~ \\
\hline
$4$ & 0 & 3 & $\tau'$ & 2.679871$_{12}$ & ~ & ~ \\
\hline
$4$ & 1 & 4 & $\tau_{\mu}$ & 2.394683$_{12}$ & ~ & ~ \\
\hline
$4$ & 2 & 1 & $\tau_{\mu\nu}$ & 1.625283$_{12}$ & 1.761046$_{28}$ & ~ \\
\hline
$4$ & 3 & 1 & $\tau_{\mu\nu\rho}$ & 1.386062$_{12}$ & 1.471897$_{24}$ & ~ \\
\hline
$4$ & 4 & 1 & $\tau_{\mu\nu\rho\sigma}$ & 1.241859$_{12}$ & 1.298227$_{24}$ & ~ \\
\hline
$5$ & 0 & 1 & $\rho$ & 2.656623$_{10}$ & ~ & ~ \\
\hline
$5$ & 0 & 3 & $\rho'$ & 3.134677$_{10}$ & ~ & ~ \\
\hline
$5$ & 2 & 1 & $\rho_{\mu\nu}$ & 2.120960$_{10}$ & ~ & ~ \\
\hline
$5$ & 3 & 1 & $\rho_{\mu\nu\rho}$ & 1.904472$_{10}$ & ~ & ~ \\
\hline
$5$ & 4 & 1 & $\rho_{\mu\nu\rho\sigma}$ & 1.724977$_{10}$ & ~ & ~ \\
\hline
$5$ & 5 & 1 & $\rho_{\mu\nu\rho\sigma\lambda}$ & 1.585858$_{10}$ & ~ & ~ \\
\hline
$6$ & 0 & 1 & $\lambda$ & 3.054421$_{10}$ & ~ & ~ \\
\hline
$6$ & 2 & 1 & $\lambda_{\mu\nu}$ & 2.555342$_{10}$ & ~ & ~ \\
\hline
$6$ & 3 & 1 & $\lambda_{\mu\nu\rho}$ & 2.357399$_{10}$ & ~ & ~ \\
\hline
$7$ & 0 & 1 & $\eta$ & 3.423800$_{10}$ & ~ & ~ \\
\hline
$7$ & 2 & 1 & $\eta_{\mu\nu}$ & 2.959249$_{10}$ & ~ & ~ \\
\hline
$7$ & 3 & 1 & $\eta_{\mu\nu\rho}$ & 2.782129$_{10}$ & ~ & ~ \\
\hline%
\end{longtable}%
}

\begin{table}[htbp]
\centering
\refstepcounter{footnote}
\caption[Critical point values $D_{\mathrm{c}}(R)$ for $V_0 = 4.0$ and $V_1 = 1.0$.]{Critical point values $D_{\mathrm{c}}(R)$, $R=\sqrt{N}$, as a function of system size $N$ for $V_0 = 4.0$ and $V_1 = 1.0$.\protect\footnotemark[\thefootnote]}%
{\small
\begin{tabular}{|c|cccccccc|}
  \hline
  $N$ & 6 & 7 & 8 & 9 & 10 & 12 &  13 &  14  \\
  \hline
  $D$ & 2.9600 & 2.9075 & 2.8747 & 2.8516  & 2.8347 & 2.8123 & 2.8054& 2.7985\\
  \hline\hline
  $N$ &  16 & 18 & 20 & 24 & 26 & 28 & 30 & 32\\ \hline
  $D$ & 2.7831 & 2.7801  & 2.7785 & 2.7756 & 2.7706 & 2.7690 & 2.7678 & 2.7677 \\
  \hline
\end{tabular}
}
\label{tab:critical_points}
\end{table}
\footnotetext[\thefootnote]{We set $V_0$ by comparing, for several candidate values, the numerically extracted scaling dimensions of a small set of levels against conformal bootstrap and exact reference data. No cost function was minimized; among the values considered, $V_0 = 4.0$ gave the closest overall agreement and is used in what follows. The overall energy scale was fixed by setting $V_1 = 1.0$.}

Figures~\ref{fig:O2SpectrumSz0}, \ref{fig:O2SpectrumSz12}, and
\ref{fig:O2SpectrumSz34}
reveal signatures of emergent conformal symmetry that include the
expected descendant structure, the stress-energy tensor, and the
conserved $SO(2)$ Noether current.  We identify 48 primary operators. For the primary
operator $\varepsilon$ corresponding to the quantum numbers
$S^{z}=0^{+}$ and $L=0$ our ED estimate is
$\Delta_{\varepsilon}=1.510065_{13}$ while our DMRG estimate is
$\Delta_{\varepsilon}=1.532935_{28}$. The subscript here and below denotes the fermion number $N$ in the ED or DMRG computation. These ED and DMRG estimates are roughly consistent with the conformal bootstrap (CB) determination of the scaling dimension $\Delta_{\varepsilon} = 1.51136(22)$~\cite{O2Bootstrap_Chester}, see
Table~\ref{tab:primaries_table}. It is a bit puzzling that the DMRG result, while referring to larger $N$, deviates from CB more than ED, as can also be seen from the dependence on $R$ of $\Delta_{\varepsilon}$ results in Fig.~\ref{fig:O2SpectrumSz0}. We stress that CB uncertainties are not directly comparable to the finite-size and numerical uncertainties in ED and DMRG. Remaining shifts at finite $R$ may reflect irrelevant operators and level mixing, as discussed for the 3D Ising model on the fuzzy sphere using conformal perturbation theory (CPT) in Ref.~\cite{Ising_CPT}.  We expect that agreement with bootstrap values will improve as $R$ is further increased, or when such CPT corrections are accounted for. The same comment applies to the comparisons below.

The exact value of the scaling
dimension $\Delta_{T_{\mu\nu}}=3$ for the stress-energy tensor
$T_{\mu\nu}$ corresponding to the lowest eigenvalue with quantum numbers $S^{z}=0^{+}$ and
$L=2$ is consistent with our ED estimate
$\Delta_{T_{\mu\nu}}=3.015277_{13}$ and DMRG estimate
$\Delta_{T_{\mu\nu}}=3.005933_{28}$ from
Table~\ref{tab:primaries_table}.  The exact value of the scaling
dimension $\Delta_{j_{\mu}}=2$ for the conserved $SO(2)$ Noether
current $j_{\mu}$ corresponding to the quantum numbers $S^{z}=0^{-}$,
$L=1$, that reflects the internal $SO(2)$ symmetry, is also
consistent with our ED estimate $\Delta_{j_{\mu}}=1.964120_{13}$ and
DMRG estimate $\Delta_{j_{\mu}}=1.987537_{28}$ from
Table~\ref{tab:primaries_table}.

The most relevant primary operator is $\sigma$, with quantum numbers $S^{z}=1$, $L=0$ and scaling dimension $\Delta_{\sigma}=0.519088(22)$ from CB~\cite{O2Bootstrap_Chester}, which plays a central role in the $O(2)$ Wilson-Fisher CFT as the analog of the order parameter in the classical $XY$ model. Since this operator is used in our determination as a reference, its ED and DMRG values in Table~\ref{tab:primaries_table} are set to the CB central value. Also appearing are irrelevant primaries, notably the leading irrelevant operator $\varepsilon'$ with ED estimate of the scaling dimension $\Delta_{\varepsilon'} = 3.887999_{13}$, which governs the leading corrections to scaling in finite-size systems. Beyond that, we find a rich spectrum of primary operators: for $S^{z} = 2$, the primary $t$ has scaling dimension $\Delta_t= 1.267871_{28}$, along with spinful primaries in that sector $t_{\mu\nu}$ ($L=2$, $\Delta= 3.023389_{28}$) and $t_{\mu\nu\rho\sigma}$ ($L=4$, $\Delta_{t_{\mu\nu\rho\sigma}}= 5.133003_{24}$);
 similarly, in the charge-3 sector, we identify the primary $\chi$ with $\Delta_{\chi}= 2.207272_{28}$, along with spinful primaries in that sector $\chi_{\mu\nu}$ ($L=2$, $\Delta= 4.018932_{28}$), $\chi_{\mu\nu\rho}$ ($L=3$, $\Delta= 4.611154_{24}$); for charges 4 and 5, the primaries $\tau$ ($\Delta = 3.316242_{28}$) and $\rho$ ($\Delta= 4.740359_{13}$) appear, respectively, along with spinful primaries in that sector. All these primaries are in rough agreement with CB and MC results, where available, see Table~\ref{tab:primaries_table}. 
 
The last column of Table~\ref{tab:primaries_table} records whether an
operator has independent support outside our calculation. We label an operator
``confirmed'' when it has previously been identified by CB or MC, or when its
scaling dimension is fixed by an exact conservation law, and ``new'' otherwise;
of the 48 primaries we identify, 26 are ``confirmed'' and 22 are ``new''. The assignments themselves rest on level counting and on the
observed finite-size trajectories: a level is called a primary only when it
cannot be consistently attached to a descendant tower already present below it.
The Supplementary Material~\cite{SuppMat} carries this out level by level for
every sector with $S^{z}\le 7$ and $L\le 5$, listing each numerical state either
as a named primary, or as a descendant at level $k$ of a given primary together
with the mismatch $\Delta_{\rm descendant}-\Delta_{\rm primary}-k$ at the largest
shared system size, or as unidentified when the available spectrum does not
support a reliable assignment. Finally, ``new'' refers only to the CB and MC
literature: many of these operators do have counterparts in the perturbative
$4-\epsilon$ expansion, and we use those matches in Sec.~\ref{sec:parity} to
assign spacetime parity.
 
 Notably, there is one primary which was previously reported by CB~\cite{chester2025bootstrappingsimplestdeconfinedquantum} although we do not see it in our data:
  \begin{itemize}
 	\item $\chi_{\mu}$: Lorentz vector and rank-3 tensor under internal $O(2)$ ($S^{z}=3$, $L=1$); $\Delta_{\chi_{\mu}}=2.078$.
 \end{itemize}
It seems unlikely that our unbiased analysis could miss such a low lying state. Also, perturbative estimates reviewed in Ref.~\cite{HENRIKSSON20231} place the lowest $S^z=3$, $L=1$ primary substantially higher. It therefore appears more plausible that the result of \cite{chester2025bootstrappingsimplestdeconfinedquantum} is a numerical artifact of the extremal functional method, although more work is needed to ascertain this.\footnote{The data in \cite{chester2025bootstrappingsimplestdeconfinedquantum} is based on the joint bootstrap analysis of the four-point function of lowest scalars of charge 0,1,2, same setup as in \cite{Liu_Simmons_Duffin_O2_spectrum}. The operator in question is extracted via the extremal functional method \cite{El-Showk:2012vjm} which is most reliable for operators with large OPE coefficients. According to \cite{Ning-private}, the extracted OPE coefficient is borderline, so a numerical artifact is a possibility. This operator is not reported in \cite{Liu_Simmons_Duffin_O2_spectrum} which did not study this sector in detail. 
	Note also that the setup of \cite{Liu_Simmons_Duffin_O2_spectrum,chester2025bootstrappingsimplestdeconfinedquantum} is not sensitive to odd-spin operators of charge 4, which explains the corresponding lacunae in the CB column of Table~\ref{tab:primaries_table}.} 

The spectrum exhibits a clear descendant structure that is
characteristic of conformal field theories. For the primary operator
$\sigma$, we observe its level-$1$ descendants at $L=1$
($\partial_{\mu}\sigma$ with $\Delta = 1+\Delta_{\sigma}$); level-$2$ descendants at $L=2$
($\partial_{\mu}\partial_{\nu}^{(0)} \sigma$
\footnote{
  The superscript $(\ell)$ indicates the angular momentum component of the descendant operator.
  }
 with $\Delta = 2+\Delta_{\sigma}$) and at $L=0$ ($\partial_{\mu}\partial_{\nu}^{(0)} \sigma$ with $\Delta = 2+\Delta_{\sigma}$), level-$3$ descendants at $L=3$
($\partial_{\mu}\partial_{\nu} \partial_{\rho}^{(3)} \sigma$ with $\Delta = 3+\Delta_{\sigma}$) and at $L=1$ ($\partial_{\mu}\partial_{\nu} \partial_{\rho}^{(1)} \sigma$ with $\Delta = 3+\Delta_{\sigma}$) and so on (see Appendix~\ref{app:primary_and_descendant_operators}). This descendant
tower structure is visible in Fig.~\ref{fig:O2SpectrumSz12} for the
$S^{z}=1$ sector, where the energy levels are organized by angular
momentum $L$.  Similar descendant structures are observed for other
primary operators, such as the descendants of $t$ in the $S^{z}=2$
sector and those of $\chi$ in the $S^{z}=3$ sector. The detailed
tables for all the levels are provided in the Supplementary Material \cite{SuppMat}.

\FloatBarrier

\subsection{Spacetime parity}
\label{sec:parity}

We close this survey of the spectrum by discussing spacetime parity, which describes how an operator changes under a reflection of space. This should not be confused with the $+$ or $-$ label in the $S^z=0$ sector: that label refers to the internal $O(2)$ symmetry. The bootstrap studies in Refs.~\cite{O2Bootstrap_Chester,Liu_Simmons_Duffin_O2_spectrum,chester2025bootstrappingsimplestdeconfinedquantum} use correlation functions of scalar operators that are even under a spatial reflection. The operators reported in these bootstrap spectra therefore have spacetime parity $P=+1$. This conclusion does not depend on whether the angular momentum $L$ is even or odd.

Information about spacetime-parity-odd operators requires a different bootstrap calculation, using correlation functions that include operators with finite angular momentum, such as the conserved current. The current bootstrap of Ref.~\cite{Dymarsky:2017xzb} and the mixed scalar-current bootstrap of Ref.~\cite{Reehorst:2019pzi} can distinguish $P=+1$ from $P=-1$. Under assumptions appropriate to the $O(2)$ model, Ref.~\cite{Reehorst:2019pzi} gives upper bounds for the first parity-odd operator in three sectors: $\Delta\lesssim 7.13$ for a neutral scalar ($S^z=0^{+}$, $L=0$), $\Delta\lesssim 8.59$ for a charge-$1$ vector ($S^z=1$, $L=1$), and $\Delta\lesssim 4.47$ for a charge-$1$ rank-$2$ Lorentz tensor ($S^z=1$, $L=2$) operator. An upper bound only says that at least one such operator must occur below the quoted value; it does not identify a particular numerical level.

All primaries marked ``confirmed'' in Table~\ref{tab:primaries_table} belong to the spacetime-parity even family, $P=+1$. Table~\ref{tab:parity_new_primaries} summarizes what can be said about the primaries marked ``new''. We first use direct matches to bootstrap spectra, and then use a perturbative match only when the charge ($S^z$), spin ($L$), ordering ($I$), and scaling dimension ($\Delta$) identify the operator unambiguously. The perturbative calculation constructs the ordinary parity-even operator family. The perturbative dimensions quoted in Table~\ref{tab:parity_new_primaries} are direct truncated $4-\epsilon$ estimates at $N=2$ and $\epsilon=1$; they are included only for unambiguous matches and have not been resummed. If neither comparison gives a unique match, we leave the parity unidentified. The level $\sigma''_{\mu\nu}$ at $\Delta=4.334637$ is a candidate for the parity-odd charge-$1$, spin-$2$ primary whose first dimension is bounded above by $4.47$ in Ref.~\cite{Reehorst:2019pzi}. The agreement is suggestive, but an upper bound alone does not prove the identification.

{\scriptsize
\begin{longtable}{|c|c|c|c|c|p{2.7cm}|}
\caption{Spacetime-parity information for primaries marked ``new'' in Table~\ref{tab:primaries_table}.}\label{tab:parity_new_primaries}\\ \hline
Operator & $(S^z,L)$ & $\Delta$ & Spacetime parity & Perturbative dimension & Basis for assignment\\
\hline\hline
$\varepsilon''$ & $(0^+,0)$ & 4.796534 & $P=+1$ & $4-\epsilon$: 4.80 & unique perturbative match \cite{HENRIKSSON20231}\\
\hline
$\varepsilon_{\mu\nu}$ & $(0^+,2)$ & 4.596819 & $P=+1$ & $4-\epsilon$: 4.61 & unique perturbative match \cite{HENRIKSSON20231}\\
\hline
$j'_\mu$ & $(0^-,1)$ & 4.156422 & $P=+1$ & $4-\epsilon$: 4.20 & unique perturbative match \cite{HENRIKSSON20231}\\
\hline
$j'_{\mu\nu\rho}$ & $(0^-,3)$ & 5.337671 & $P=+1$ & $4-\epsilon$: 5.37 & unique perturbative match \cite{HENRIKSSON20231}\\
\hline
$\sigma''_{\mu\nu}$ & $(1,2)$ & 4.334637 & candidate for $P=-1$ & -- & compatible with the mixed-bootstrap bound \cite{Reehorst:2019pzi}\\
\hline
$t'_{\mu\nu}$ & $(2,2)$ & 4.680865 & $P=+1$ & $4-\epsilon$: 4.47 & unique perturbative match \cite{HENRIKSSON20231}\\
\hline
$t_{\mu\nu\rho}$ & $(2,3)$ & 5.350336 & $P=+1$ & $4-\epsilon$: 5.42 & unique perturbative match \cite{HENRIKSSON20231}\\
\hline
$\chi'$ & $(3,0)$ & 5.131775 & $P=+1$ & $4-\epsilon$: 5.30 & unique perturbative match \cite{HENRIKSSON20231}\\
\hline
$\chi_\mu$ & $(3,1)$ & 5.590976 & $P=+1$ & $4-\epsilon$: 5.30 & unique perturbative match \cite{HENRIKSSON20231}\\
\hline
$\chi_{\mu\nu\rho\sigma\lambda}$ & $(3,5)$ & 6.620238 & $P=+1$ & $4-\epsilon$: 6.65 & unique perturbative match \cite{HENRIKSSON20231}\\
\hline
$\tau'$ & $(4,0)$ & 6.616791 & $P=+1$ & $4-\epsilon$: 7.00 & unique perturbative match \cite{HENRIKSSON20231}\\
\hline
$\tau_\mu$ & $(4,1)$ & 6.894136 & unidentified & -- & no unique match\\
\hline
$\tau_{\mu\nu\rho}$ & $(4,3)$ & 5.932580 & $P=+1$ & $4-\epsilon$: 5.62 & unique perturbative match \cite{HENRIKSSON20231}\\
\hline
$\rho'$ & $(5,0)$ & 8.236994 & unidentified & -- & no unique match\\
\hline
$\rho_{\mu\nu}$ & $(5,2)$ & 6.684267 & unidentified & -- & no unique match\\
\hline
$\rho_{\mu\nu\rho}$ & $(5,3)$ & 7.367956 & unidentified & -- & no unique match\\
\hline
$\rho_{\mu\nu\rho\sigma}$ & $(5,4)$ & 7.831272 & unidentified & -- & no unique match\\
\hline
$\rho_{\mu\nu\rho\sigma\lambda}$ & $(5,5)$ & 8.136111 & unidentified & -- & no unique match\\
\hline
$\lambda_{\mu\nu}$ & $(6,2)$ & 8.149670 & unidentified & -- & no unique match\\
\hline
$\lambda_{\mu\nu\rho}$ & $(6,3)$ & 8.878337 & unidentified & -- & no unique match\\
\hline
$\eta_{\mu\nu}$ & $(7,2)$ & 9.716965 & unidentified & -- & no unique match\\
\hline
$\eta_{\mu\nu\rho}$ & $(7,3)$ & 10.391125 & unidentified & -- & no unique match\\
\hline%
\end{longtable}%
}

\subsection{OPE Coefficients}
\label{sec:ope_coefficients}

We extract the diagonal OPE coefficients
$f_{\mathrm{o}\varepsilon\mathrm{o}}(R)$
in the scaling ansatz (\ref{eq:deltaE_o close to criticality})
with the help of
\begin{equation}
f_{\mathrm{o}\varepsilon\mathrm{o}}(R)=
\left.
\frac{\partial\,\delta\,E_{\mathrm{o}}(R,D)}{\partial\,g_{\varepsilon}(R, D)}
\right|^{\,}_{g_{\varepsilon}(R, D)=0}.
\label{eq:def fovarepsilono(R)}
\end{equation}
This expression follows from first-order conformal perturbation theory,
where the energy shift due to the relevant operator $\varepsilon$
is proportional to the OPE coefficient $f_{\mathrm{o}\varepsilon\mathrm{o}}$
that describes how the operator $\mathrm{o}$ fuses with
$\varepsilon$ to produce itself.
For primary operators, $f_{\mathrm{o}\varepsilon\mathrm{o}}$ is a universal
OPE coefficient, while for descendants it is related
to the primary OPE coefficient
through known functions of the scaling dimensions%
~\cite{Icosahedron,Ising_CPT}.
Throughout this subsection, the term ``OPE coefficient'' refers to $f_{\mathrm{o}\varepsilon\mathrm{o}}$.

\begin{figure}[htbp]
  \centering
  \includegraphics[width=\textwidth]{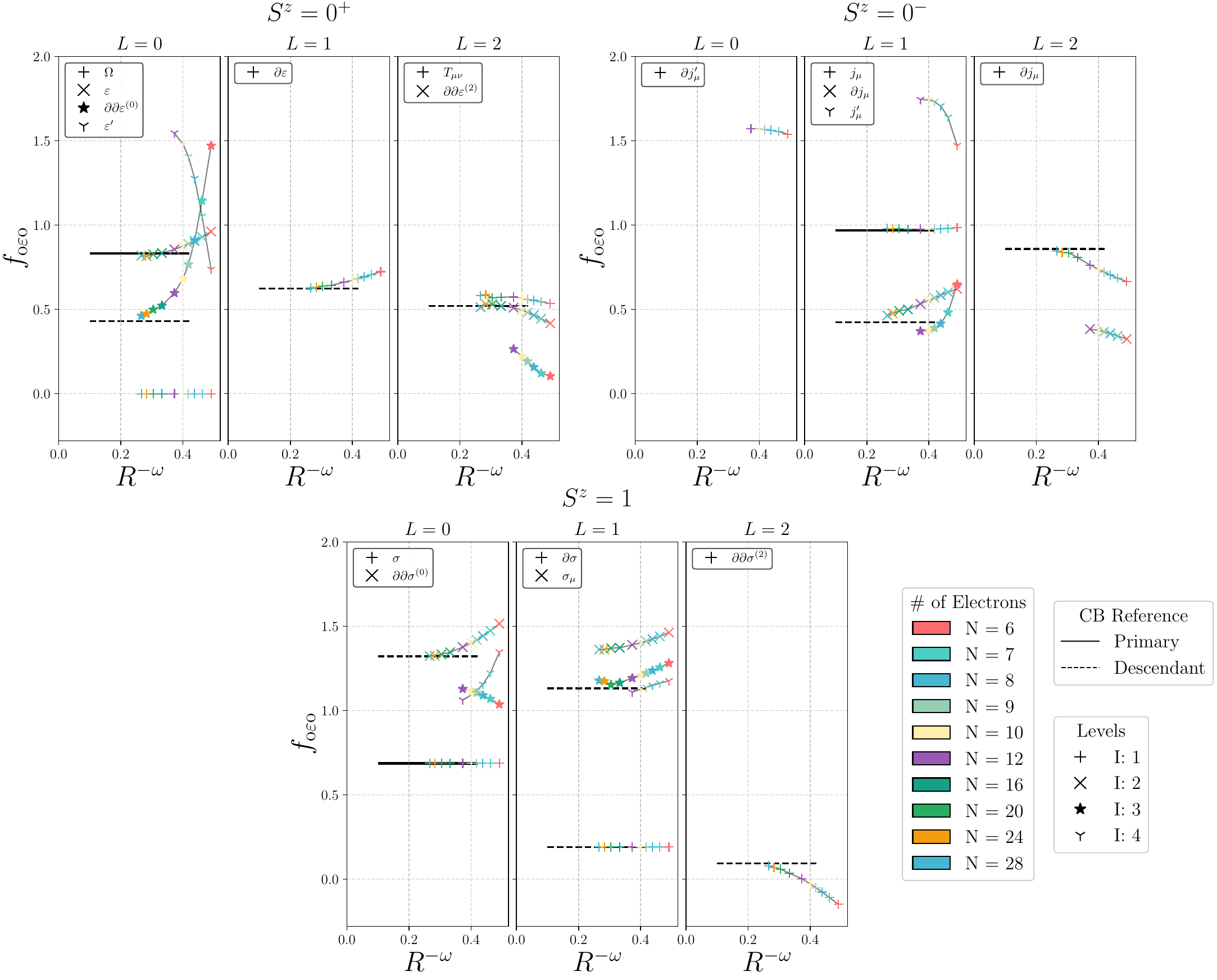}
  \caption{Dependence on $R^{-\omega}$ with $\omega\equiv\Delta_{\varepsilon'}-3$
of the OPE coefficient
$f_{\mathrm{o}\varepsilon\mathrm{o}}(R)$
defined by Eq.\ (\ref{eq:def fovarepsilono(R)})
for operators in the $S^{z}=0^{+}$, $S^{z}=0^{-}$, and $S^{z}=1$ sectors
for the projection
(\ref{eq:symmetry group quantum fuzzy Hamiltonian e})
of the total spin along the quantization axis,
organized by 
the label $L$ of the Casimir operator for the total
orbital angular momentum
(\ref{eq:symmetry group quantum fuzzy Hamiltonian b}). The Level label $I$ has the same meaning as in Tables~\ref{tab:primaries_table} and \ref{tab:opes_with_primaries}.
\label{fig:OPE_combined}
}
\end{figure}

In \cfig{OPE_combined},
we plot the dependence on $R^{-\omega}$ of
$f_{\mathrm{o}\varepsilon\mathrm{o}}(R)$
for selected OPE coefficients across different $SO(2)$ charge sectors,
where $\omega\equiv\Delta_{\varepsilon'}-3$ is the anomalous dimension
of the leading irrelevant operator.
The limiting values as $R\uparrow\infty$ give the universal OPE coefficients
$f_{\mathrm{o}\varepsilon\mathrm{o}}$ in the (2+1)D $O(2)$ Wilson-Fisher CFT.
Our estimates for these OPE coefficients, obtained from the largest available
ED and DMRG system sizes, are reported in Table~\ref{tab:opes_with_primaries}.

The top-left panel of Fig.~\ref{fig:OPE_combined} ($S^{z}=0^{+}$) arranges curves by $L=0$, $1$, and $2$.
In the $L=0$ column the first level above the vacuum is $\varepsilon$; our $f_{\varepsilon\varepsilon\varepsilon}=0.818711_{28}$ lies close to the CB determination $0.830914(32^{*})$, and the $R^{-\omega}$ extrapolation approaches it systematically.
The next displayed level is the second descendant of $\varepsilon$; its coefficient trends toward the primary CB determination multiplied by the descendant factor (cf. Appendix~\ref{App: CPT}).
The fourth level is $\varepsilon'$; the corresponding OPE coefficient is not reported in the CB literature;
at twelve electrons (our largest size for that level) the flow in $R^{-\omega}$ still drifts rather than settling.
The $L=1$ column shows the first descendant of $\varepsilon$. The $L=2$ column shows $T_{\mu\nu}$, the spin-$2$ second descendant of $\varepsilon$, and the first descendant of $T_{\mu\nu}$. The $\varepsilon$ descendants approach the CB value multiplied by the corresponding descendant factors, while the stress-tensor trajectories have no CB reference lines in this plot.
The top-right panel ($S^{z}=0^{-}$) shows $f_{j_{\mu}\varepsilon j_{\mu}}(R)$ nearly flat in $R^{-\omega}$, with modest finite-size corrections.
Descendants of $j_{\mu}$ converge more slowly, and several trajectories remain short of their asymptotes at the sizes available here.
We also resolve the primary $j_{\mu}'$. The scalar level in the $L=0$ column is identified as its level-$1$ descendant; both trajectories still show significant finite-size drift.
In the bottom panel ($S^{z}=1$), the ED/DMRG results for $\sigma$ and $\partial\sigma$ coincide with CB identically, as expected because those levels calibrate the critical point and the speed of light.
The displayed second descendants $\partial\partial\sigma^{(0)}$ and $\partial\partial\sigma^{(2)}$ trend toward the limits expected from the descendant relations. The $L=1$ primary $\sigma_{\mu}$ has no CB reference line in this plot and retains visible finite-size drift.

For a spinful primary $\mathrm{o}$ with $L\neq 0$, the three-point function $\langle \mathrm{o}\varepsilon \mathrm{o}\rangle$
may admit several independent conformally invariant tensor structures, hence more than one OPE coefficient; the quantity $f_{\mathrm{o}\varepsilon\mathrm{o}}$ extracted here is fixed by Hamiltonian state normalization and is then a linear combination of those coefficients, to be identified case by case.\footnote{See App.~B.4 of \cite{Ising_CPT} where the corresponding linear combination was dubbed $f^{\rm shift}$.}
In particular, $f_{T_{\mu\nu}\varepsilon T_{\mu\nu}}=0.581426_{28}$ for the stress tensor and $f_{j_{\mu}\varepsilon j_{\mu}}=0.979486_{28}$ for the conserved $O(2)$ Noether current.\footnote{In the latter case there is in fact only one independent OPE coefficient, and ours agrees with the CB-extracted one once normalization convention is properly taken into account, see App.~\ref{sec:J}.} 
For primaries with nonzero $U(1)$ charge\footnote{Strictly speaking, for charged primaries the relevant three-point function is $\langle \mathrm{o}^*\varepsilon \mathrm{o}\rangle$, and the corresponding OPE coefficient is $f_{\mathrm{o}^*\varepsilon\mathrm{o}}$.} and with $L=0$, namely $t$, $\chi$, and $\tau$, we find $f_{t\varepsilon t}=1.255077_{24}$, $f_{\chi\varepsilon\chi}=1.770280_{28}$, and $f_{\tau\varepsilon\tau}=2.244990_{28}$.
Together with the preceding results, these estimates further characterize the operator algebra; for descendants, the extracted coefficients agree with the primary expectations up to the descendant factors of Appendix~\ref{App: CPT}.

\subsection{Large-Charge Expansion}
\label{sec:large_charge}

The large-charge expansion%
~\cite{Hellerman15,Monin17,Alvarez-Gaume17,Cuomo2020,Gaume21}
provides a powerful semiclassical framework for understanding
operators in a CFT with an internal continuous symmetry
when these operators carry a large charge $Q$ of this internal continuous
symmetry.
For a CFT in $(2+1)$ dimensions,
the lowest primary operator carrying
large charge $Q$ is predicted to be a scalar and its scaling dimension obeys the expansion%
~\cite{Hellerman15,Monin17,Cuomo2020}
\begin{equation}
\Delta_Q = \alpha Q^{3/2} + \beta Q^{1/2} - 0.0937256 + O(Q^{-1/2}).
\label{eq:large_charge_expansion}
\end{equation}
The leading $Q^{3/2}$ term arises from the classical energy of a homogeneous charge configuration on a two-sphere of radius $R$,
for which the semiclassical scaling dimension behaves as $\Delta\sim Q^{3/2}$ in $(2+1)$ dimensions.
The subleading $Q^{1/2}$ term and the constant encode quantum fluctuations of the Goldstone mode (including the one-loop Casimir contribution). The $\alpha$ and $\beta$ are Wilson coefficients that
characterize the effective field theory whose values at present cannot be computed analytically; they have been estimated from Monte Carlo simulations \cite{Banerjee:2017fcx, Cuomo:2023rjg,hasenbusch2025precisionestimateslargecharge}.\footnote{In what follows we will mostly compare with the most recent \cite{hasenbusch2025precisionestimateslargecharge} which has the highest accuracy for the observables we need here.}
 On the other hand the constant term in \eqref{eq:large_charge_expansion} is computable as a one-loop determinant.\footnote{This was first pointed out in \cite{Hellerman15}, with an error in the numerical value due to incorrect one-loop determinant regularization. The correct value was first given in \cite{Monin17}.}  Its value was also checked by Monte Carlo simulations \cite{hasenbusch2025precisionestimateslargecharge}.

In our context,
the internal continuous symmetry is the $U(1)\cong SO(2)$ symmetry that is
generated by the projection of the total spin operator
along the quantization axis as given by Eq.\
(\ref{eq:symmetry group quantum fuzzy Hamiltonian e}),
i.e., the large charge is the eigenvalue $S^{z}$ of $\widehat{S}^{z}$. The Wilson coefficients $\alpha$, $\beta$ in
Eq.~(\ref{eq:large_charge_expansion}) must be determined from
independent data. In our analysis, we use the CB central values for scaling dimensions of
the lowest primaries at $S^{z} = 2$ and $S^{z} = 4$
to fix these coefficients $\alpha$, $\beta$, which gives:
\begin{equation}
\alpha = 0.33094,\quad \beta = 0.27856.
\label{eq:ouralphabeta}
\end{equation}
This compares well with the previous Monte Carlo determination $\alpha = 0.33163(9)$, $\beta = 0.2765(8)$ \cite{hasenbusch2025precisionestimateslargecharge}.

 In the fit yielding \eqref{eq:ouralphabeta}, we set to zero the $O(Q^{-1/2})$ correction in Eq.~(\ref{eq:large_charge_expansion}). This assumption may appear questionable, since the used values of the charge are not large by any measure. However, when we then use the large-charge expansion formula \eqref{eq:delta_with_phonons} to predict the lowest $S^z=3$ primary dimension, we get a value close to the CB result. This check partly vindicates our fitting procedure.

In each sector of fixed large charge $S^{z}$, the large-charge effective theory on the sphere describes a conformal superfluid ground state, which is the lowest CFT primary with charge $S^z$. Phonon excitation modes above this ground state describe other low-lying CFT states in the charge  $S^z$ sector. While phonon excitations would be gapless in infinite volume, on a sphere of radius $R$ the phonon spectrum is discrete. The phonon frequencies are \cite{Hellerman15}
\begin{equation}
\omega_L = \frac1 R\sqrt{\frac{L(L+1)}{2}},
\qquad
L=1,2,3,\ldots.
\label{eq:phonon_dispersion}
\end{equation}
Here $L$ labels angular momentum on the sphere. In what follows we set $R=1$.
Adding $L=1$ phonons, for which $\omega_L=1$, does not yield new CFT primaries but descendants of the primary corresponding to the ground state. On the other hand, adding phonons with $L \geq 2$ yields new primaries in the charge $S_z$ sector.

Hence, the scaling dimensions of low-lying primary operators in the charge-$S^{z}$ sector are predicted to be
\begin{equation}
\Delta = \Delta_{S^{z}} + \sum_{L \geq 2} n_L\, \omega_L\,,
\label{eq:delta_with_phonons}
\end{equation}
where $\Delta_{S^{z}}$ is the scaling dimension of the lowest primary in the charge-$S^{z}$ sector, obtained from \eqref{eq:large_charge_expansion} with $Q=S^{z}$.
The integers $n_L\in\{0,1,2,\ldots\}$ are occupation numbers of phonons of angular momentum $L\ge 2$, and $\omega_L$ is given by \eqref{eq:phonon_dispersion}.
States with all $n_L=0$ are large-charge ground states.

We refer to excitations with $n_L\neq 0$ as phonon primaries, and we match the predictions of \eqref{eq:delta_with_phonons} to our numerical data to identify these operators at the $(2+1)$D $O(2)$ Wilson-Fisher quantum critical point. In particular, we identify the phonon primaries with $n_L=1$ (and all other $n_{L'}=0$) for a given $L$. The corresponding primary has spin $L$.

In our numerical results (Figs.~\ref{fig:O2SpectrumSz12} and
\ref{fig:O2SpectrumSz34}), the cyan lines show the predictions of \eqref{eq:delta_with_phonons} in each $S^{z}$ sector for
the ground states ($n_L=0$ for all $L$) and for phonon primaries with $n_L=1$ for a given $L\ge 2$.
No cyan line is drawn for the lowest primaries at $S^{z}=2$ and $S^{z}=4$, because their CB scaling dimensions are inputs that fix $\alpha$ and $\beta$ (see \eqref{eq:ouralphabeta}).
The figures show the large-charge ground states in the $L=0$ sectors; spectra at higher $S^{z}$ appear in the Supplemental Material~\cite{SuppMat}.

\begin{figure}[t]
    \centering
    \includegraphics[width=\textwidth]{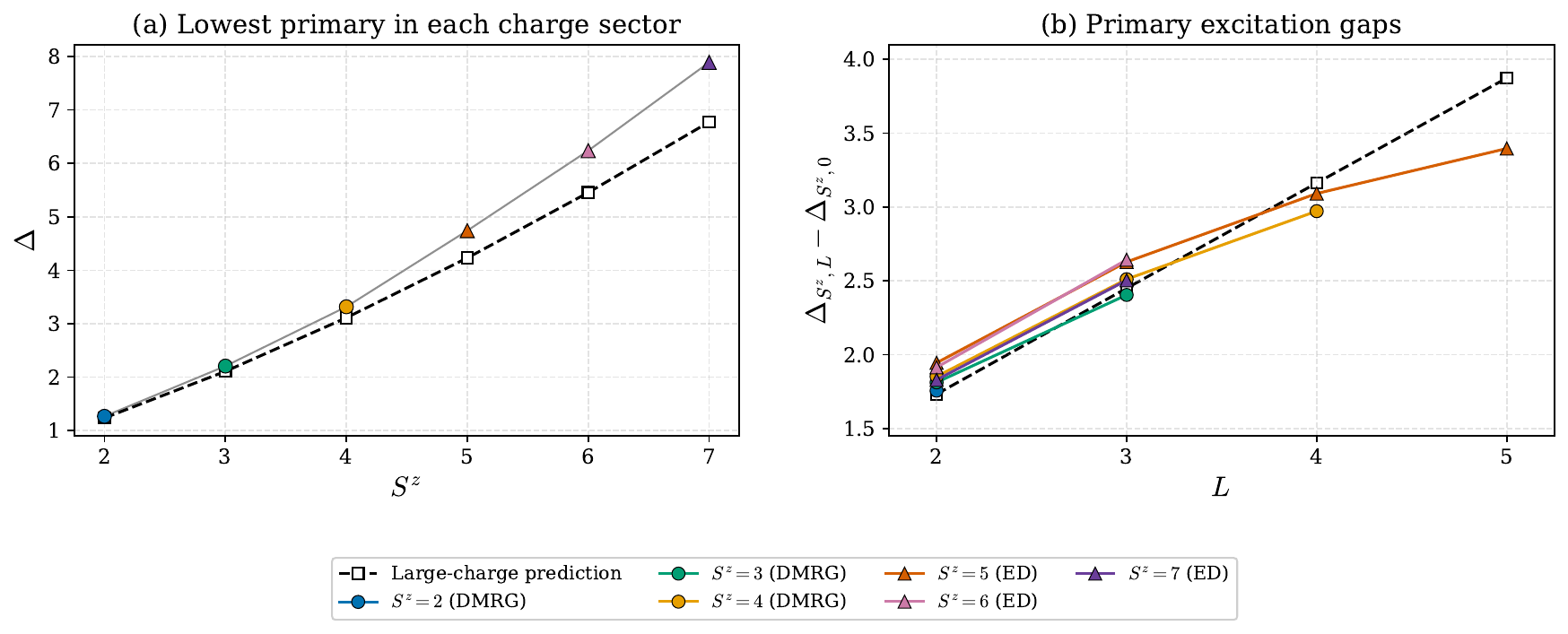}
    \caption{Comparison of the large-charge result with the largest available ED or DMRG value. Panel (a) shows the lowest $L=0$ primary in each charge sector. Panel (b) shows the dimensions of the primaries assigned to the one-phonon sequence relative to the lowest primary in the same charge sector. The dashed black line is the prediction $\omega_L$ of Eq.~\eqref{eq:phonon_dispersion}. Only reliably assigned levels are included; for $S^z=6,7$ these extend through $L=3$. Circles denote DMRG data and triangles ED data.\label{fig:large_charge_comparison}}
\end{figure}

The comparison is summarized in Fig.~\ref{fig:large_charge_comparison}. The identified one-phonon sequences occur at $L\leq S^z$. For $S^{z}=2$, $t_{\mu\nu}$ at $L=2$ follows the one-phonon prediction. The primaries $t_{\mu\nu\rho}$ and $t_{\mu\nu\rho\sigma}$ at $L=3,4$ lie above the corresponding predictions and are not assigned to this sequence.
For $S^{z}=3$, $\chi_{\mu\nu}$ and $\chi_{\mu\nu\rho}$ at $L=2,3$ follow the one-phonon prediction.
For $S^{z}=4$, we observe $\tau_{\mu\nu}$, $\tau_{\mu\nu\rho}$, and $\tau_{\mu\nu\rho\sigma}$ at $L=2$, $L=3$, and $L=4$.
For $S^{z}=5$, the sequence continues with $\rho_{\mu\nu}$, $\rho_{\mu\nu\rho}$, $\rho_{\mu\nu\rho\sigma}$, and $\rho_{\mu\nu\rho\sigma\lambda}$ at $L=2,3,4,5$. For $S^{z}=6$ and $S^{z}=7$, we identify the $L=2,3$ pairs $\lambda_{\mu\nu}$, $\lambda_{\mu\nu\rho}$ and $\eta_{\mu\nu}$, $\eta_{\mu\nu\rho}$, respectively; the higher-spin levels remain unidentified. Panel (a) shows a common upward finite-size shift in the absolute dimensions at these higher charges. Subtracting the numerical ground-state dimension in panel (b) removes most of this shift: the $L=2,3$ gaps, as well as the $S^{z}=5$, $L=4$ gap, lie close to the one-phonon prediction. The $S^{z}=5$, $L=5$ gap shows a larger deviation. Although its absolute dimension, $8.136111_{13}$, is close to the large-charge value $8.102$.
We also identify an $L=5$ primary $\chi_{\mu\nu\rho\sigma\lambda}$ at $S^{z}=3$. Its value $\Delta=6.782848_{24}$ is well above the one-phonon prediction $5.981$, so we do not assign it to the phonon sequence.

Theoretically, it has been argued \cite{Cuomo:2017vzg,Cuomo2020} that, for large charge, the phonon excitation formula \eqref{eq:delta_with_phonons} should apply up to spin $L\lesssim \sqrt{Q}$ ($Q=S^z$) while for $\sqrt{Q}\lesssim L \lesssim Q $ the lowest excitation is a vortex anti-vortex pair, and $\Delta$ grows logarithmically with $L$. For even higher $Q\lesssim L \lesssim Q^{3/2}$ the lowest excitations were argued to be described by multiple vortex-antivortex pairs \cite{Cuomo:2017vzg,Cuomo2020}  while for $Q^{3/2}\lesssim L \lesssim Q^{2}$ by a ``giant vortex'' \cite{Cuomo_2023}. 

Our results provide some of the first numerical checks of these theoretical conclusions.
~\footnote{
\AD{In Ref.~\cite{chester2025bootstrappingsimplestdeconfinedquantum}, phonon primaries were similarly compared to CB data, but the phonon primary at $S^{z}=4$, $L=3$ was not observed.}} 
We saw that the simple phonon excitation formula fits the spectrum well up to $L=S^z$, but not beyond that, already for $S^z=O(1)$, which may be surprising. We did not see any crossover from the phonon to vortex-antivortex behavior at $L=\sqrt{S^z}$. The same observation was previously made in \cite{chester2025bootstrappingsimplestdeconfinedquantum}(App.D) based on the analysis of the CB data. This issue requires further study.

For $S^z=2,3,4,5$, the absolute dimension of the identified $L=S^z$ primary lies close to the large-charge prediction. The excitation-gap comparison in Fig.~\ref{fig:large_charge_comparison}, however, shows a larger deviation for the $S^z=5$, $L=5$ state, so the absolute agreement should not be overinterpreted.

Physically, we would like to think of the phonon primaries as the critical point counterparts of the Goldstone phonons from the long-range-ordered superfluid phase. 
This interpretation applies only to the subset of operators we identify as phonon primaries through \eqref{eq:delta_with_phonons}; we do not assign a comparable characterization to the remainder of the critical spectrum. Nevertheless, this example shows how operators at a conformal fixed point may connect to excitations familiar from other gapless phases.

\section{Conclusion}
\label{sec:conclusion}

We have presented a comprehensive numerical study of the
$O(2)$ Wilson-Fisher conformal field theory
in $(2+1)$-dimensional spacetime
using a fuzzy-sphere regularization.
Employing exact diagonalization for
systems up to $N = 13$ and density matrix renormalization group for
systems up to $N = 28$, we identified 48 primary operators across the
charge sectors $S^{z} = 0$ through $S^{z} = 7$. The extracted scaling
dimensions show good agreement with CB and MC results, where available, see Table~\ref{tab:primaries_table}, and additionally we find several primaries that are not discussed in the CB or MC literature.
We also extracted OPE coefficients using
conformal perturbation theory and verified predictions from the large
charge expansion, demonstrating the connection between the existence of
a single Goldstone (phonon) mode
in the superfluid phase and phonon primaries at criticality.

Several natural extensions of this work present themselves.
This study extracts only
diagonal OPE coefficients
$f_{\mathrm{o}\varepsilon\mathrm{o}}$ from the linear response to the
relevant perturbation. Computing four-point correlation functions on
the fuzzy sphere, as demonstrated for the Ising CFT
in $(2+1)$-dimensional spacetime
\cite{Fuzzy_4pt_correlators},
would give access to the full set of
off-diagonal OPE coefficients and provide a direct test of crossing
symmetry in the $O(2)$ theory. The Lorentzian inversion formula
applied to $O(2)$ bootstrap data~\cite{Liu_Simmons_Duffin_O2_spectrum}
yields predictions for an extensive set of OPE data in various charge
sectors, against which such fuzzy-sphere four-point functions could be
quantitatively compared. Constructing the infrared conformal
generators on the fuzzy sphere~\cite{Fuzzy_Generators, Fan2024, fardelli2026improving3disingope}, already
achieved for the Ising case, would further sharpen the distinction
between primaries and descendants in the $O(2)$ spectrum and enable
the systematic extraction of higher-point conformal data.

A natural question raised by this work is how far the precision of
fuzzy-sphere conformal data can be pushed. Our finite-size estimates
for different operators converge at markedly different rates, with the
stress tensor reaching $\Delta_{T_{\mu\nu}} \approx 3.0$ at modest system sizes,
while other operators exhibit larger finite-size drifts. Since these
corrections are governed by the coupling to irrelevant operators in
the conformal perturbation theory framework, one could in principle
tune the microscopic Hamiltonian, for instance by adjusting the
pseudopotential parameters or including higher angular momentum
channels, to suppress the leading irrelevant contributions without
changing the universality class. Such an optimization, combined with
quantum Monte Carlo methods as recently formulated on the fuzzy
sphere~\cite{Fuzzy_QMC_Ising} (which can access significantly larger
system sizes than ED or DMRG)
could bring the fuzzy-sphere approach
closer to the precision of the conformal bootstrap for the $O(2)$
theory.

Improving this precision is not merely a technical exercise. The
critical exponents of the $O(2)$ universality class
in $(2+1)$-dimensional spacetime
are among the most precisely measured quantities in all of physics, thanks to
microgravity experiments on the specific heat of superfluid helium-4
near the lambda point~\cite{Lipa2003}. A long-standing tension
persists between these experimental measurements and the theoretical
values obtained from the conformal
bootstrap~\cite{O2Bootstrap_Chester} and Monte Carlo simulations~\cite{Hasenbusch_2019_improvedclockmodel}, with
the correlation length exponent $\nu$ showing a significant discrepancy of $8\sigma$~\cite{JCCM_January_2020_02}. Independent numerical approaches with distinct systematic
uncertainties are therefore valuable. The fuzzy sphere, as a method
that directly accesses the operator spectrum rather than thermodynamic
quantities, offers a complementary perspective on this discrepancy,
and its continued development may help clarify whether the
disagreement points to unaccounted experimental systematics or to gaps
in our theoretical understanding of one of the most fundamental phase
transitions in nature.

\appendix
\section{Haldane Pseudopotentials}
\label{app:pseudopotentials}

The two-body interaction matrix elements $V^{\,}_{m_1,m_2,m_3,m_4}$ in
Eq.~(\ref{eq:H_00}) and (\ref{eq:H_xy}) are constructed using Haldane
pseudopotentials~\cite{Haldane_1983_FQHE}, which parametrize two-body
$SO(3)$-invariant interactions on the fuzzy sphere. The matrix
elements are given by
\begin{equation}
  V^{\,}_{m_1,m_2,m_3,m_4} =
  \sum_{l=0}^{2s} V^{\,}_l
  \sum_{M=-l}^{l}
  \langle m_1, m_2 | l, M \rangle \langle l, M | m_3, m_4 \rangle,
\end{equation}
where $\langle m_1, m_2 | l, M \rangle$ are Clebsch-Gordan
coefficients coupling two single-particle states with magnetic quantum
numbers $m_1$, $m_2$ to a two-particle state with relative angular
momentum $l$ and $z$-component $M = m_1 + m_2$. Angular momentum
conservation requires $m_1 + m_2 = m_3 + m_4$. The pseudopotential
$V^{\,}_l$ controls the interaction strength in the relative angular
momentum $l$ channel. For our calculations, we restrict to $l \leq 1$,
so only $V^{\,}_{0}$ and $V^{\,}_{1}$ are nonzero.

The pseudopotentials provide a systematic decomposition of
rotationally invariant two-body interactions on the sphere into
channels of definite relative angular
momentum~\cite{Haldane_1983_FQHE,Zhu_Han_Huffman_Hofmann_He_2023}.
Using Wigner 3j-symbols, the matrix elements can equivalently
be written as
\begin{equation}
  V^{\,}_{m_1,m_2,m_3,m_4} =
  \sum_{l=0}^{2s} V^{\,}_l\,(4s-2l+1)
  \begin{pmatrix} s & s & 2s-l \\ m_1 & m_2 & -m_1-m_2 \end{pmatrix}
  \begin{pmatrix} s & s & 2s-l \\ m_3 & m_4 & -m_3-m_4 \end{pmatrix},
  \label{eq:pseudopotential_3j}
\end{equation}
where the 3j-symbols couple two spin-$s$ particles to total angular
momentum $2s-l$ (i.e., relative angular momentum~$l$).

In real space on the sphere, the
pseudopotential $V^{\,}_l$ corresponds to the LLL projection of a
specific central interaction.
$V^{\,}_{0}$ is the LLL projection of the contact interaction
$\delta^{(2)}(\bm{n}-\bm{n}')$,
where $\bm{n},\bm{n}'$ are unit vectors on $S^{2}$:
when two particles sit at the same point,
their relative angular momentum vanishes ($l=0$).
$V^{\,}_{1}$ is the LLL projection of the gradient interaction
$\nabla^{2}\delta^{(2)}(\bm{n}-\bm{n}')$,
which probes the $l=1$ channel.
Restricting to $l\leq 1$ thus retains only the two
most short-range components of the interaction.

\FloatBarrier
\section{Exact Diagonalization Implementation}
\label{app:ED}

This appendix provides details for reproducing our exact
diagonalization (ED) calculations. The
Hamiltonian~(\ref{eq:H_fuzzy_sphere}) describes spin-1 fermions on the
fuzzy sphere. For a system with monopole strength $s$, there are $N =
2s+1$ orbitals per spin orientation, and with three spin states
($\sigma = 0, \pm 1$), the total number of single-particle orbitals is
$3N$. We work at one-third filling with $N_e = N$ electrons. For example,
our largest ED system size $N = 13$ corresponds to 13 electrons
distributed among 39 single-particle orbitals.

The Hamiltonian preserves three symmetries enabling
block-diagonalization: (1) spatial $SO(3)$ rotation symmetry,
conserving total angular momentum $L$ and its $z$-component $L_z$; (2)
internal $SO(2)$ spin rotation about the $z$-axis, conserving $S^{z}$ as in Eq.~(\ref{eq:symmetry group quantum fuzzy Hamiltonian e});
and (3) $\mathbb{Z}_2$ parity in the $S^{z} = 0$ sector, corresponding
to $c_{m,\sigma} \to c_{m,-\sigma}$. We construct symmetry-resolved
Hilbert spaces by generating all Fock basis states $\prod_{i=1}^{N_e}
\hat{c}^{\dagger}_{m_{i},\sigma_{i}} \ket{0}$ for fixed particle
number $N_e$, total spin $S^{z}$, and center-of-mass orbital momentum
$\bar{m} = \sum_{i} m_{i}$. In contrast to periodic boundary
conditions on the torus, the orbital momenta $m_{i}$ are not folded
into a Brillouin zone.

Hamiltonian matrix elements are computed in the Fock basis. The
interaction matrix elements $V^{\,}_{m_1,m_2,m_3,m_4}$ are obtained
from Haldane pseudopotentials and Wigner 3j-symbols (see
Appendix~\ref{app:pseudopotentials}). We restrict to $l \leq 1$ in the
pseudopotential expansion, so only $V^{\,}_{0}$ and $V^{\,}_{1}$ are
nonzero. To find the low-energy spectrum, we employ the implicitly
restarted Arnoldi iteration as implemented in
ARPACK~\cite{Lehoucq1998arpack}, which requires only matrix-vector
products to compute extremal eigenvalues. The largest
symmetry-resolved Hilbert space dimension occurs in the $L_z = 0$,
$S^{z} = 1$ sector with approximately $42.3 \times 10^6$ basis states
at $N = 13$.

\section{Density Matrix Renormalization Group Implementation}
\label{app:DMRG}

The Density Matrix Renormalization Group (DMRG)
method~\cite{White1992,Schollwock2011DMRG} enables access to larger
system sizes than ED by variationally optimizing a matrix product
state (MPS) ansatz. We use ITensor~\cite{ITensor} as the underlying
tensor network library.

The three spin-1 degrees of freedom ($\sigma = 0, \pm 1$) at each
orbital are represented as a chain of $3N$ spinless fermion sites,
ordered by ascending orbital quantum number $m$. The MPS bond
dimension $\chi$ controls the entanglement that can be captured; we
use bond dimensions up to $\chi = 3072$ to reach system sizes of $N =
28$ electrons in 84 orbitals.

We implement the following Abelian symmetries to reduce computational
cost and target specific quantum number sectors: (1) $U(1)$
conservation of total particle number $N_e$; (2) $U(1)$ conservation
of $L_z$, the $z$-component of angular momentum; and (3) $U(1)$
conservation of $S^{z}$, the total spin along the $z$-axis.

The matrix product operator (MPO) representation of the Hamiltonian is
constructed from the two-body interaction matrix elements
$V^{\,}_{m_1,m_2,m_3,m_4}$. Terms with amplitude below $10^{-12}$ are
discarded, and the MPO is compressed using standard techniques.

We employ two-site DMRG with noise~\cite{White2005} to ensure
ergodicity in the variational optimization, using a noise amplitude of
$10^{-5}$. Convergence is monitored through variations in the ground
state energy and mid-chain entanglement entropy, with a threshold of
$10^{-7}$. To compute excited states, we add weighted projectors onto
previously computed low-energy states to the effective Hamiltonian.

To obtain states with definite angular momentum $L$, we use an
orthogonalization procedure starting from the highest $L_z$
sector. For a target $L$, we first compute eigenstates in the $L_z =
L$ sector. When computing states in lower $L_z$ sectors, we
orthogonalize against all previously computed states from higher
angular momentum multiplets. This ensures that states obtained at $L_z
= \ell$ with no component in higher $L_z$ sectors belong to the $L =
\ell$ multiplet. As a consistency check, we compute the expectation
value of $L^{2}$ for each state to verify the correct angular momentum
assignment. The $\mathbb{Z}_2$ parity symmetry is not implemented
directly in DMRG; instead, we determine the parity of $S^{z} = 0$
states by explicit evaluation after the calculation.

\section{Primary and Descendant Operators}
\label{app:primary_and_descendant_operators}

Conformal field theories are characterized by the structure of their
operator content, which organizes into representations of the
conformal group $SO(d+1,1)$. The state-operator correspondence on the
sphere $S^{d-1} \times \mathbb{R}$ provides a direct link between
eigenstates of the Hamiltonian and CFT
operators~\cite{Di_Francesco_Mathieu_Senechal_1997,Rychkov_2017, Zhu_Han_Huffman_Hofmann_He_2023}. Specifically,
eigenstates of a quantum Hamiltonian defined on the sphere $S^{d-1}$
are in one-to-one correspondence with scaling operators of the
infrared CFT, and the energy gaps are proportional to the scaling
dimensions of their corresponding operators.

The operators in a CFT are organized into conformal multiplets, each
consisting of a primary operator and its descendants. A primary
operator $\mathrm{o}$ is an operator that is annihilated by the
special conformal generators at the origin, and it is characterized by
its scaling dimension $\Delta_{\mathrm{o}}$ and its transformation
properties under the rotation group $SO(d)$ (for $d=3$, this is
$SO(3)$ which corresponds to angular momentum $L$ on the sphere
$S^{2}$). Descendants are generated by acting on the primary with
momentum operators (derivatives: $\partial_{\mu}$) $P_{\mu}$, which increase the energy
by exactly 1 unit per level.

On $S^{2} \times \mathbb{R}$ (which corresponds to our
fuzzy-sphere regularization), the descendant structure is
particularly transparent. For a scalar primary operator with angular
momentum $L_P = 0$ and scaling dimension $\Delta$, the level-$n$
descendants have energy $\Delta + n$ and angular momenta $L_D \in \{n,
n-2, n-4, \cdots\}$ down to 0 (if $n$ is even) or 1 (if $n$ is
odd). For example, the level-1 descendant $\partial_{\mu}\mathrm{o}$
has angular momentum $L_D = 1$, while level-2 descendants
$\partial_{\mu}\partial_{\nu}\mathrm{o}$ decompose into a symmetric
traceless tensor with $L_D = 2$ and a scalar trace part with $L_D =
0$.

For spinning primaries (i.e., primaries with $L_P > 0$), the
descendant structure is more complex. At level $n$, the allowed
angular momenta are $L_D \in \{L_P + n, L_P + n - 1, L_P + n - 2,
\cdots, max(0,|L_P - n|)\}$.

Conserved quantities, such as the stress-energy tensor $T_{\mu\nu}$
and conserved currents like the $SO(2)$ Noether current $j_{\mu}$, have
special descendant structures due to their conservation laws. The
stress-energy tensor has scaling dimension $\Delta = d$ (3 in our
case) and angular momentum $L_P = 2$, while conservation
$\partial_{\mu}T_{\mu\nu} = 0$ eliminates certain descendant states,
leading to shortened multiplets. Similarly, conserved currents have
$\Delta = d-1$ (2 in our case) and their conservation eliminates the
scalar descendant at level 1. For such cases, there is no descendant with $L_D < L_P$.

\section{Conformal Perturbation Theory}
\label{App: CPT}

Conformal perturbation theory provides a systematic framework to
understand finite-size corrections to the CFT spectrum on $\mathbb{R}
\times S^{2}$~\cite{Icosahedron}. Any regularization of the sphere
introduces deviations from the exact CFT, which can be captured by
perturbing the CFT Hamiltonian:
\begin{align}
	\widehat{H}=
	\widehat{H}^{\,}_{\text{CFT}}
	+
	\sum_{i} g_{i} \int_{S^{2}} \mathcal{V}_{i},
\end{align}
where $\mathcal{V}_{i}$ are local CFT operators and $g_{i}$ are
effective couplings that depend on the microscopic details. At first
order in perturbation theory, the energy correction for a state
$|\psi\rangle$ corresponding to operator $\mathrm{o}$ is
\begin{align}
	\delta E_{\mathrm{o}} = 4\pi g_\varepsilon f_{\mathrm{o}\varepsilon\mathrm{o}},
\end{align}
where $f_{\mathrm{o}\varepsilon\mathrm{o}}$ is the OPE coefficient
and $g_\varepsilon$ is the coupling to the perturbing operator
$\varepsilon$.

For descendants of a scalar primary $\mathrm{o}$, the OPE
coefficients $f_{\mathrm{o}\varepsilon\mathrm{o}}$ in
Eq.~(\ref{eq:deltaE_o close to criticality}) are related to those of
the primary through universal factors determined by conformal
symmetry~\cite{Icosahedron,Ising_CPT}. For the level-$k$ descendant in
irreducible representation $\rho$ of $SO(3)$:
\begin{align}
	f_{\partial^k\mathrm{o},\varepsilon,\partial^k\mathrm{o}}^{(\rho)} =
	f_{\mathrm{o}\varepsilon\mathrm{o}} \, A(k,\rho),
\end{align}
where the factor $A(k,\rho)$ depends on $\Delta_{\mathrm{o}}$ and
$\Delta_{\varepsilon}$. For the first descendant ($k=1$, $\rho=3$):
\begin{align}
	A(1,3) = 1 + \frac{\Delta_{\varepsilon}(\Delta_{\varepsilon}-3)}{6\Delta_{\mathrm{o}}}.
	\label{eq:descendant_factor}
\end{align}
The perturbing operator enters through its conformal Casimir
eigenvalue $C_{\varepsilon} =
\Delta_{\varepsilon}(\Delta_{\varepsilon}-3)$, with $A(k,\rho) = 1$
for marginal perturbations ($\Delta_{\varepsilon}=3$). Higher-level
factors $A(k,\rho)$ for $k \leq 4$ were given in \cite{Icosahedron}, see the Mathematica notebook \cite{CPTnotebook} for a full list of requisite results and derivations. They were used in Ref.~\cite{Ising_CPT}.

\section{Noether Current OPE Coefficient and Descendant Factors}
\label{sec:J}

The Noether current OPE coefficient $JJS$ was determined in a bootstrap study \cite{Reehorst:2019pzi} as (see their Eq.~(3))
\begin{equation}
|\lambda_{JJS}| = 0.645(4)
\end{equation}
Note our $j_\mu,\varepsilon$ are their $J,S$. They could only determine $\lambda_{JJS}$ up to a sign. The coefficient $f_{j_\mu \varepsilon j_\mu}$ is related to $\lambda_{JJS}$ by a rescaling:
\begin{equation}
	f_{j_\mu \varepsilon j_\mu} = \frac 23 \Delta_\varepsilon (3-\Delta_\varepsilon) \lambda_{JJS},
	\label{eq:Jresc}
\end{equation}
which gives CB $f_{j_\mu \varepsilon j_\mu}$ value reported in Table \ref{tab:opes_with_primaries}.

Let's retrace the main steps giving \eqref{eq:Jresc}. The ``standard'' normalization of the 3pt function conserved vector -- conserved vector -- parity-even
scalar is given by 
\begin{equation}
	\begin{array}{lll}
		\langle J (P_1, Z_1) J (P_2, Z_2) S (P_3) \rangle (P_{12})^{2 -
			\frac{\Delta_S}{2}} (P_{13})^{\Delta_S / 2} (P_{23})^{\Delta_S / 2} & = &
		\tilde{\lambda}_{J J S} \left( H_{12} + \frac{\Delta_S}{\Delta_S - 2} V_{1, 23} V_{2, 31} \right)
	\end{array} \label{l-mine}
\end{equation}
where we put $d=3$, $\lambda^{(2)} = \tilde{\lambda}_{J J \varphi}$
in Eq.~(2.4) ($\ell=0$) of \cite{Dymarsky:2017xzb} and set $\lambda^{(1)} = \frac{\Delta_S}{\Delta_S - 2}\lambda^{(2)}$ as enforced by the conservation of the vector. In this case there is only one independent OPE coefficient consistent with the conservation. The two-point function is assumed unit-normalized, 
		$\langle J J\rangle = \frac{H_{12}}{P_{12}^{2}}$.
	The embedding space conventions for $H_{ij}$ and $V_{i,jk}$ in Eq.~(2.3) of \cite{Dymarsky:2017xzb} differ from the original conventions of \cite{Costa:2011mg} by denominators. 
	The OPE coefficient $f_{j_\mu S j_\mu}$ is related (for $d=3$) to $\tilde{\lambda}_{J J S}$ via (see below)
	\begin{equation}
		f_{J J S} = \frac{2 (\Delta_S - 3)}{3 (\Delta_S - 2)} \tilde{\lambda}_{J J
		S}\,. \label{f-l}
	\end{equation}
This is in our convention in which for the CPT perturbing Hamiltonian $\delta \hat H = g_{S}
	\int_{S^2} S(x)$, the energy shift of the $J$ state is $4\pi f_{J J S}\,  g_S $.
	
Ref.~\cite{Reehorst:2019pzi} normalized the two and three point functions as\footnote{We thank Emilio Trevisani for communications helping us to clarify their normalization conventions.} (see their App.~A)
\begin{equation}
	\langle J J \rangle = \frac{C_J}{(4 \pi)^2} \frac{\hat{H}_{12}}{x_{12}^4},\qquad 
	\langle J J S \rangle = \frac{C_J}{(4 \pi)^2} \hat{\lambda} \frac{(\Delta_S
	- 2) \hat{H}_{12} + \Delta_S \hat{V}_{1, 23} \hat{V}_{2, 31}}{| x_{12}
	|^{4-\Delta_S} | x_{13} |^{\Delta_S} | x_{23} |^{\Delta_S}}
	\end{equation}
with $\hat{\lambda} = - \Delta_S \lambda_{J J S}$, where the structures $\hat H$ and $\hat V$ are the physical space projections of $H$ and $V$, by 
\begin{equation}
	P_{i j} \rightarrow x_{i j}^2, Z_1 Z_2 \rightarrow z_1 z_2, P_1 Z_2
	\rightarrow z_2 x_{12}\,.
\end{equation}
Comparing with \eqref{l-mine} (and not forgetting to rescale $J$ to have the
unit 2pt function) we have $\tilde{\lambda}_{JJS} = \hat{\lambda} (\Delta_S - 2)$. Finally using \eqref{f-l} we obtain \eqref{eq:Jresc}.

To get \eqref{f-l} one needs to take the 3pt function \eqref{l-mine} in position space, send one current to 0, one to infinity, and integrate the scalar insertion over the sphere. This computes the energy shift of the current state from the conformal perturbation by the scalar. This computation can be found in \cite{CPTnotebook}, where the relative energy shifts for the level one descendants are also computed. 

\acknowledgments

AML and AD acknowledge helpful discussions with G.~Cuomo, Y.-C.~He, J.~Penedones, and R.~Rattazzi. SR thanks David Poland and Emilio Trevisani for useful communications.
LH acknowledges the Tremplin funding from CNRS Physique and was also supported by the ANR JCJC ANR-25-CE30-2205-01. SR~is partially supported by the Simons Collaboration on the Probabilistic Paths to Quantum Field Theory (award SFI-MPS-PP-00012621-16).

\bibliographystyle{JHEP}
\bibliography{bibtex.bib}
\end{document}


\maketitle
\flushbottom

\section{Scaling dimensions data tables}
\label{sec:delta_tables}

For each level we give a plausible but conservative multiplet identification based on level counting and the observed finite-size trajectories. A level is called a primary only when it cannot be consistently assigned to the descendant towers already present below it. When an assignment is not reliable, the level is left unidentified. For a descendant, the defect is $\Delta_{\rm descendant}-\Delta_{\rm primary}-k$, evaluated at the largest system size shared by the two levels.

\FloatBarrier
\inserttable{0}{+}{0}
\inserttable{0}{+}{1}
\inserttable{0}{+}{2}
\inserttable{0}{+}{3}
\inserttable{0}{+}{4}
\inserttable{0}{+}{5}

\FloatBarrier
\inserttable{0}{-}{0}
\inserttable{0}{-}{1}
\inserttable{0}{-}{2}
\inserttable{0}{-}{3}
\inserttable{0}{-}{4}
\inserttable{0}{-}{5}

\FloatBarrier
\inserttable{1}{}{0}
\inserttable{1}{}{1}
\inserttable{1}{}{2}
\inserttable{1}{}{3}
\inserttable{1}{}{4}
\inserttable{1}{}{5}

\FloatBarrier
\inserttable{2}{}{0}
\inserttable{2}{}{1}
\inserttable{2}{}{2}
\inserttable{2}{}{3}
\inserttable{2}{}{4}
\inserttable{2}{}{5}

\FloatBarrier
\inserttable{3}{}{0}
\inserttable{3}{}{1}
\inserttable{3}{}{2}
\inserttable{3}{}{3}
\inserttable{3}{}{4}    
\inserttable{3}{}{5}

\FloatBarrier
\inserttable{4}{}{0}
\inserttable{4}{}{1}
\inserttable{4}{}{2}
\inserttable{4}{}{3}
\inserttable{4}{}{4}
\inserttable{4}{}{5}

\FloatBarrier
\inserttable{5}{}{0}
\inserttable{5}{}{1}
\inserttable{5}{}{2}
\inserttable{5}{}{3}
\inserttable{5}{}{4}
\inserttable{5}{}{5}

\FloatBarrier
\inserttable{6}{}{0}
\inserttable{6}{}{1}
\inserttable{6}{}{2}
\inserttable{6}{}{3}
\inserttable{6}{}{4}
\inserttable{6}{}{5}

\FloatBarrier
\inserttable{7}{}{0}
\inserttable{7}{}{1}
\inserttable{7}{}{2}
\inserttable{7}{}{3}
\inserttable{7}{}{4}
\inserttable{7}{}{5}

\FloatBarrier
\section{Operator product expansion coefficients data tables}
\label{sec:ope_tables}

Each table below lists, for one symmetry sector, the OPE coefficient
$f_{\mathrm{o}\varepsilon\mathrm{o}}$ measured at every available system size,
next to two columns that say what the level is and what is expected of it.

The ``Identification'' column simply repeats the assignment made for the
same level in Section~\ref{sec:delta_tables}.

The ``Comments'' column answers two different questions, depending on the
level. The first concerns descendants. A descendant does not carry an OPE
coefficient of its own: conformal symmetry fixes it to be that of its primary
multiplied by a pure number, $f_{\rm desc}=A\,f_{\rm prim}$. The factor $A$
depends on how many derivatives were applied, on the spin of the level they
produce, and on the dimensions of the primary and of $\varepsilon$. We therefore
quote $A$ together with the prediction $A\,f_{\rm prim}$, where $f_{\rm prim}$ is
our own value for the primary read off at the largest system size the two levels
share; comparing that prediction with the measured numbers in the same row is a
direct check of the assignment. These factors are known for descendants of
scalar primaries up to the fourth level, and for the conserved current and the
stress tensor at the low levels computed in Ref.~\cite{CPTnotebook}. Where no
factor is available the entry is left as ``--''.

The second question concerns primaries that carry spin, and is a warning
about what the tabulated number means. For a scalar primary the three-point
function with $\varepsilon$ is fixed by a single number, so the measured $f$ is
that number. For a primary of spin $L$ this is no longer true: the three-point
function is built from $L+1$ independent invariants, with coefficients
$\lambda_{0},\dots,\lambda_{L}$, and the energy shift on the sphere responds to
only one fixed combination of them, denoted $f^{\rm shift}$ in
Ref.~\cite{Ising_CPT}. That combination is what the comment records, for instance
$f\propto\lambda_{1}-\tfrac{1}{3}\lambda_{0}$ for $L=1$; its coefficients are
pure numbers, independent of the operator dimensions. The individual $\lambda_{i}$ cannot be recovered from our data, so
these entries are a statement about interpretation rather than a prediction. For
the conserved current and the stress tensor the $\lambda_{i}$ are related to one
another by conservation and only one of them is independent.

\FloatBarrier
\insertopetable{0}{+}{0}
\insertopetable{0}{+}{1}
\insertopetable{0}{+}{2}
\insertopetable{0}{+}{3}
\insertopetable{0}{+}{4}
\insertopetable{0}{+}{5}

\insertopetable{0}{-}{0}
\insertopetable{0}{-}{1}
\insertopetable{0}{-}{2}
\insertopetable{0}{-}{3}
\insertopetable{0}{-}{4}
\insertopetable{0}{-}{5}

\FloatBarrier
\insertopetable{1}{}{0}
\insertopetable{1}{}{1}
\insertopetable{1}{}{2}
\insertopetable{1}{}{3}
\insertopetable{1}{}{4}
\insertopetable{1}{}{5}

\FloatBarrier
\insertopetable{2}{}{0}
\insertopetable{2}{}{1}
\insertopetable{2}{}{2}
\insertopetable{2}{}{3}
\insertopetable{2}{}{4}
\insertopetable{2}{}{5}

\FloatBarrier
\insertopetable{3}{}{0}
\insertopetable{3}{}{1}
\insertopetable{3}{}{2}
\insertopetable{3}{}{3}
\insertopetable{3}{}{4}
\insertopetable{3}{}{5}

\FloatBarrier
\insertopetable{4}{}{0}
\insertopetable{4}{}{1}
\insertopetable{4}{}{2}
\insertopetable{4}{}{3}
\insertopetable{4}{}{4}
\insertopetable{4}{}{5}

\FloatBarrier
\insertopetable{5}{}{0}
\insertopetable{5}{}{1}
\insertopetable{5}{}{2}
\insertopetable{5}{}{3}
\insertopetable{5}{}{4}
\insertopetable{5}{}{5}

\FloatBarrier
\insertopetable{6}{}{0}
\insertopetable{6}{}{1}
\insertopetable{6}{}{2}
\insertopetable{6}{}{3}
\insertopetable{6}{}{4}
\insertopetable{6}{}{5}

\FloatBarrier
\insertopetable{7}{}{0}
\insertopetable{7}{}{1}
\insertopetable{7}{}{2}
\insertopetable{7}{}{3}
\insertopetable{7}{}{4}
\insertopetable{7}{}{5}

\FloatBarrier
\clearpage

\bibliographystyle{JHEP}
\bibliography{bibtex.bib}